%% file: parallel_surgery_PRXQresub.tex
\documentclass[a4paper, onecolumn, 11pt, unpublished]{quantumarticle}
\pdfoutput=1
\usepackage{breakurl}
\usepackage{stmaryrd}
\usepackage{amsthm,amstext,amssymb,amsmath}
\usepackage{epsfig}
\usepackage{slashed}
\usepackage{fancyhdr}
\usepackage{graphicx}
\usepackage{pdfpages}
\usepackage{float}
\usepackage{simplewick}
\usepackage{graphics,psfrag}
\usepackage{pstricks}
\usepackage{latexsym}
\usepackage{changepage}
\usepackage{caption}
\usepackage{paralist}
\usepackage{tikzit}
\input{basic.tikzstyles}
\input{quantum.tikzstyles}
\usepackage{graphicx}
\usepackage{array}
\usepackage{hyperref}
\usepackage{tikz-cd}
\usepackage{mathrsfs}
\usepackage{braket}
\usepackage{soul}
\usepackage{pifont}
\usepackage{subcaption}

\usepackage{hhline}
\usepackage{algorithm}
\usepackage{algpseudocode}

\usepackage{scalefnt}
\usepackage[all]{xypic}
\usepackage{slashed}
\usepackage{extarrows}
\usepackage{lettrine}
\usepackage{booktabs}
\usepackage{paralist}
\usepackage{wrapfig}
\usepackage{xcolor}

\usepackage{thmtools}

\DeclareMathAlphabet{\mathcal}{OMS}{cmsy}{m}{n}
\newtheorem{lemma}{Lemma}[section] 

\newtheorem{corollary}[lemma]{Corollary}
\newtheorem{example}[lemma]{Example}
\newtheorem{theorem}[lemma]{Theorem}

\newtheorem{definition}[lemma]{Definition}
\newtheorem{remark}[lemma]{Remark}

\renewcommand{\imath}{\mathrm{i}}

\newcommand{\CC}{\hbox{{$\mathcal C$}}}
\newcommand{\CA}{\hbox{{$\mathcal A$}}}
\newcommand{\CF}{\hbox{{$\mathcal F$}}}
\newcommand{\CG}{\hbox{{$\mathcal G$}}}

\newcommand{\CV}{\hbox{{$\mathcal V$}}}
\newcommand{\CH}{\hbox{{$\mathcal H$}}}
\newcommand{\CS}{\hbox{{$\mathcal S$}}}

\newcommand{\CL}{\hbox{{$\mathcal L$}}}

\newcommand{\CP}{\hbox{{$\mathcal P$}}}
\newcommand{\CQ}{\hbox{{$\mathcal Q$}}}

\newcommand{\CE}{\hbox{{$\mathcal E$}}}
\newcommand{\CO}{\hbox{{$\mathcal O$}}}

\newcommand{\CI}{\hbox{{$\mathcal I$}}}

\newcommand{\C}{\mathbb{C}}

\newcommand{\F}{\mathbb{F}}
\newcommand{\FF}{\mathbb{F}}

\newcommand{\del}{\partial}

\newcommand{\im}{\mathrm{im}}
\newcommand{\<}{\langle}
\renewcommand{\>}{\rangle}

\newsavebox{\pullback}
\sbox\pullback{%
\begin{tikzpicture}%
\draw (0,0) -- (2ex,0ex);%
\draw (2ex,0ex) -- (2ex,2ex);%
\end{tikzpicture}}

\newsavebox{\pushout}
\sbox\pushout{%
\begin{tikzpicture}%
\draw (0,0) -- (0,2ex);%
\draw (0,2ex) -- (2ex,2ex);%
\end{tikzpicture}}

\setstcolor{red}

\newcommand{\xmark}{\ding{55}}

\usepackage{cite}

\begin{document}

\title{Parallel Logical Measurements via Quantum Code Surgery}


\author{Alexander Cowtan}
\affiliation{Department of Computer Science, University of Oxford, Oxford OX1 3QD, UK}
\email{akcowtan@gmail.com}
\author{Zhiyang He}
\affiliation{Department of Mathematics, Massachusetts Institute of Technology, Cambridge, MA 02139, USA}
\author{Dominic J. Williamson}
\affiliation{IBM Quantum, IBM Almaden Research Center, San Jose, CA 95120, USA}
\author{Theodore J. Yoder}
\affiliation{IBM Quantum, IBM T.J. Watson Research Center, Yorktown Heights, NY 10598, USA}

\maketitle

\begin{abstract}
    Quantum code surgery is a flexible and low-overhead technique for performing logical measurements on quantum error-correcting codes, which generalises lattice surgery. 
    In this work, we present a code surgery scheme, applicable to any qubit stabiliser low-density parity check (LDPC) code, that fault-tolerantly measures many logical Pauli operators in parallel. 
    For a collection of logically disjoint Pauli product measurements supported on $t$ logical qubits, our scheme uses $\mathcal{O}\big(t \omega (\log t + \log^3\omega)\big)$ ancilla qubits, where $\omega \geq d$ is the maximum weight of the single logical Pauli representatives involved in the measurements, and $d$ is the code distance.
    This is all done in time $\mathcal{O}(d)$ independent of $t$.
    Our proposed scheme preserves both the LDPC property and the fault-distance of the original code, without requiring ancillary logical codeblocks which may be costly to prepare.
    This addresses a shortcoming of several recently introduced surgery schemes which can only be applied to measure a limited number of logical operators in parallel
    if they overlap on data qubits.
\end{abstract}

\section{Introduction}

Topological codes, specifically surface codes, have long been the leading candidates for the practical implementation of large-scale fault-tolerant quantum computation. 
Recently, experiments have demonstrated sub-threshold scaling of surface codes~\cite{google2023suppressing,acharya2024quantumerrorcorrectionsurface} and logical operations on 2D color codes~\cite{bluvstein2024logical,rodriguez2024experimental,lacroix2024scaling}. 
Despite their many desirable properties, topological codes in flat space have a notable limitation: their low encoding rate leads to a significant space overhead in achieving fault tolerance.
This limitation motivates the study of quantum Low-Density Parity Check (LDPC) codes (see Ref.~\cite{BE2} for a recent review), which have higher rate and therefore serve as candidates for low-overhead fault-tolerant quantum memories~\cite{BCGMRY,XBAP}. 

For topological codes, particularly surface codes~\cite{Kit}, methods for logical computation are well-studied~\cite{HFDM,moussa2016transversal,gidney2024magic}.
Quantum LDPC codes, however, require new techniques to perform logical computation. 
Several methods have been proposed, such as constructing codes with native transversal gates~\cite{grassl2013leveraging,Burt1,quintavalle2023partitioning,ES,sayginel2024fault,malcolm2025computing,zhu2023non,scruby2024quantum,golowich2024quantum,lin2024transversal,hsin2024classifying,breuckmann2024cups,zhu2025topological}, homomorphic logical measurements~\cite{HJY, XZ}, and generalised surgery~\cite{Coh}. 
Transversal gates can in principle be performed in constant time~\cite{De,ZZ} and with no additional overhead on the quantum code given a sufficiently powerful decoder, but cannot be universal by the Eastin-Knill theorem~\cite{EK}, and also require a quantum code to have particular structure.\footnote{See also introduction of Ref.~\cite{he2025quantum} for a discussion on the challenge of performing addressable logic with transversal gates.}
Homomorphic measurements and generalised code surgery (also known as LDPC code surgery or simply code surgery) are both techniques that perform logical Pauli measurements on logical qubits, which makes them well-suited to Pauli-based computation~\cite{BSS} where Pauli measurements and magic states together enable universal quantum computation.\footnote{In Ref.~\cite{litinski2019game}, for instance, a circuit made of Clifford and T gates is compiled into Pauli measurements with ancillary T magic states, rendering the application of logical Clifford gates unnecessary, in principle.}
Nonetheless, homomorphic logical measurements are, thus far, restricted to topological codes and homological product codes. 
By contrast, generalised code surgery is applicable to any LDPC code.\footnote{Another subtle difference between generalised code surgery and homomorphic measurement is that surgery has more flexible control over which set of logical qubits one wishes to address, while homomorphic measurement is, prior to this work, more parallel than surgery.}

To perform code surgery, we initialise an ancilla system, consisting of new qubits and stabiliser generators, to deform the quantum code in such a way that chosen logical operators are decomposed into stabilisers, a procedure that can be viewed as gauge-fixing or weight reduction~\cite{VLCABT,hastings2016weight,hastings2021quantum,Coh,Sabo2024,WY,IGND}. Code surgery performs a logical measurement on the decomposed operators, which can be made fault-tolerant by measuring all of the new stabiliser generators for a sufficient number of rounds (typically taken to be the code distance $d$) to prevent timelike errors.
After obtaining the logical measurement outcome, we then measure out the ancillary qubits and return to the original code.

In this work, we present a code surgery scheme that performs parallel measurements of logical Pauli operators, applicable to any LDPC codes. For a collection of logically disjoint Pauli product measurements (see Def.~\ref{def:logical_disjoint} and Fig.~\ref{fig:parallel_circuit}) supported on $t$ logical qubits, our scheme uses ${\mathcal{O}\big(t \omega (\log t + \log^3\omega)\big)}$ ancilla qubits, where $\omega \geq d$ is the maximum weight of the single logical Pauli representatives involved in the measurements.

\begin{figure}[t]\centering
\input{tikz_files/parallel_circuit.tikz}
\caption{An example logical circuit with three timesteps of Pauli product measurements. Each timestep contains logically disjoint Pauli products, meaning that they act on different logical qubits, see Def.~\ref{def:logical_disjoint}. The logically disjoint Pauli products are measured in parallel.
}\label{fig:parallel_circuit}
\end{figure}

This provides access to a wide array of parallel logical operations which can be complemented by transversal magic gates or magic state preparation to achieve universal fault-tolerant quantum computation.

\subsection{Related literature}
In this work we build upon substantial recent improvements to quantum LDPC code surgery, which we non-exhaustively review here, eliding many subtleties. 
Quantum LDPC code surgery was first investigated by Cohen et al.~\cite{Coh} using Tanner graphs, with a space overhead of $\CO(\omega d) \geq \CO(d^2)$ per logical qubit to be measured, given logical operator representatives with weight $\omega$. 
This space overhead was required to guarantee fault tolerance. 
We call this the CKBB scheme from now on.
Cowtan \& Burton~\cite{CowBu} took a different approach, using homology, yielding a similar construction where fault tolerance is not guaranteed but the overhead is reduced. 
This was found to be sufficient to preserve the distance in small practical cases~\cite{Cow24}.
Shortly after Ref.~\cite{Cow24}, the independent work of Cross et al.~\cite{CHRY} showed that if the measured logical operators have suitable expansion properties, then the overhead of the CKBB scheme can be qualitatively reduced without a guarantee that the deformed code is LDPC.
Later, Ide et al.~\cite{IGND} and Williamson \& Yoder~\cite{WY} independently proposed that one can directly construct an ancilla system with sufficient expansion (instead of relying on the existing expansion properties of the operators) to measure an arbitrary Pauli logical operator with space overhead $O(\omega\log^3 \omega)$, where the $\log^3\omega$ factor comes from graph decongestion~\cite{FH} and the deformed code remains LDPC.

\begin{remark}
    More precisely, the decongestion overhead was not discussed in Ref.~\cite{IGND}, beyond stating that it is polylogarithmic, as they found that in practice it was not required for the codes of interest. In the first arXiv preprint version of Ref.~\cite{WY} it was erroneously claimed that the decongestion overhead was $\CO(\log^2\omega)$.
\end{remark}
Performing Pauli product measurements of disjoint logical operators was further improved by using adapters to connect their ancilla systems in Swaroop et al.~\cite{SJY}, building on earlier ideas of bridging from Ref.~\cite{CHRY} and applying the gauging measurement directly to disjoint logical operators in Ref.~\cite{WY}. 
In summary, current methods require $\CO(r\omega\log^3\omega)$ ancillary qubits to measure the product of $r$ logical Pauli operators that are disjoint on the physical data qubits.

Unfortunately, while there have been vast improvements to the original $\CO(d\omega)$ overhead of the CKBB scheme, parallelisation has presented a persistent obstacle. 
This obstacle is present because, generically, the logical operators to be measured in a high-rate LDPC code overlap substantially on data qubits as in Figure~\ref{fig:summary}(b), with the overlap growing with the rate and distance of the code. Consequently, deforming a code without drastically increasing stabiliser weights or stabiliser overlap, and thereby losing the LDPC property, is challenging. 


To address the parallelisation challenge, Zhang \& Li~\cite{ZL} proposed techniques called brute-force branching and devised sticking, such that logical operators of the same type, say $\bar{Z}$ Pauli product measurements, can be measured in parallel using a single ancilla system. At a high level, brute-force branching uses \textit{branching stickers} to create new, disjoint representatives of logical operators on ancilla qubits, which are then measured using the CKBB scheme. Devised sticking, on the other hand, measures all logical operators in a region of the code at once using the CKBB scheme, but requires additional checks and qubits to avoid measuring unwanted logical operators.

Zhang \& Li's techniques have three main limitations: (a) they use the CKBB scheme for measurement, which leads to worst-case space overheads of $\CO(t\omega(\log t + d))$ and $\CO(t^2\omega d)$ for measurement by brute-force branching and devised sticking respectively, both of which are at least quadratic in $d$ as $\omega \geq d$ \footnote{Note that even if $\omega \in \CO(n)$, $d$ will be asymptotically much larger than $\log^3\omega$ unless $d$ is exponentially smaller than $n$. Thus when taking the same logical representatives as input, our scheme is asymptotically more efficient than Zhang \& Li in most reasonable cases, with hyperbolic surface codes a notable exception~\cite{breuckmann2017hyperbolic}.}. (b) Their scheme cannot measure operators containing $Y$ terms in parallel without using ancillary logical $\ket{Y}$ states, as proposed by Ref.~\cite[Fig.~10]{CC} in the context of twist-free lattice surgery. 
To prepare these logical $\ket{Y}$ states, Zhang \& Li required that the LDPC code used must support a transversal logical $S$ gate which is a non-trivial assumption. Distillation, as an alternative, would be much more costly. Moreover, even to measure operators containing a mixture of $X$ and $Z$ terms in parallel Zhang \& Li required ancillary logical $\ket{0}$ states. (c) Their scheme is specific to Calderbank-Shor-Steane (CSS) codes, rather than more general stabiliser codes.

\subsection{Contributions}\label{sec:contributions}

Our surgery scheme features two important improvements. 
First of all, we use the gauging measurement scheme of Ref.~\cite{WY} instead of the CKBB scheme, which leads to a significantly lower space overhead and more rigorous guarantees on fault tolerance.
Secondly, we consider the problem of measuring a logically disjoint collection of Pauli product operators, defined as follows.
\begin{definition}\label{def:logical_disjoint}
    For a quantum stabiliser code, suppose we fixed a basis of $X$-type and $Z$-type logical operators $\bar{X}_1, \cdots, \bar{X}_k$ and $\bar{Z}_1, \cdots, \bar{Z}_k$ for the $k$ logical qubits, thereby fixing $Y$-type logical operators $\bar{X}_1\bar{Z}_1,\cdots, \bar{X}_k\bar{Z}_k$ up to scalars. 
    A logical Pauli product operator $P$ is simply a product of these single logical qubit operators, up to multiplication by stabilisers. 
    The logical support of $P$ is the set of logical qubits on which $P$ acts non-trivially.
    A collection of logical Pauli product operators $P_1, \cdots, P_\ell$ is logically disjoint if their logical supports are disjoint.
\end{definition}

Intuitively, logically disjoint Pauli product operators act on different \textit{logical} qubits, i.e. there is no overlap in the logical qubit support.  For example, $\bar{X}_1\otimes \bar{Z}_2$ and $\bar{Z}_3\otimes \bar{X}_4$, as shown in the first timestep of Fig.~\ref{fig:parallel_circuit}, are logically disjoint, but $\bar{X}_1\otimes \bar{Z}_2$ and $\bar{Z}_2\otimes \bar{Z}_3$, shown in timesteps 1 and 2 respectively, are not. Logically disjoint Pauli product operators need not act on different \textit{physical} qubits, and the physical qubit supports may overlap substantially.

Our scheme can measure logically disjoint mixed type operators that contain logical $Y$ terms in parallel without using ancilla $\ket{Y}$ states or codeblocks. This is a stronger result than the scheme for measuring operators of the same type ($X$ or $Z$) introduced in Ref.~\cite{ZL}. In the same vein, our scheme can also measure logical operators in parallel on stabiliser codes which are not CSS.

We combine several elements from the recent literature to obtain our results. 
First, we use a particular basis \cite[Sec.~4.1]{gottesman1997stabilizer} and a generalised version of brute-force branching from Ref.~\cite{ZL} to obtain copies of the logicals to be measured.
Then we construct ancillary expander graphs for each measurement, following Refs.~\cite{WY,IGND}, and use the adapters from Ref.~\cite{SJY} to connect the ancillary systems to generate product measurements.

\begin{figure}
    \centering
    \includegraphics[width=0.85\linewidth]{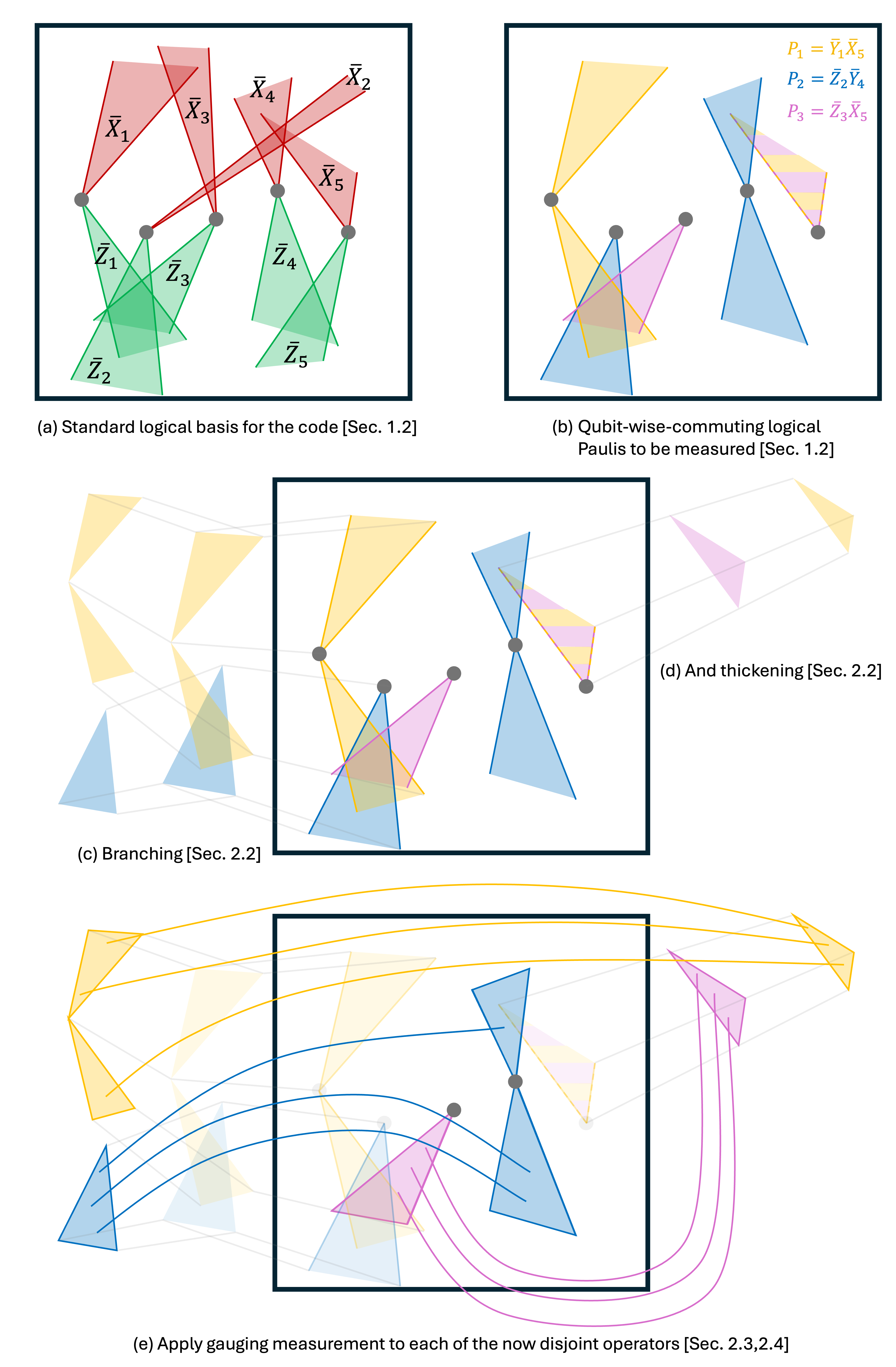}
    \caption{A summary of our core construction. (a) We start with an arbitrary quantum LDPC code and a standard logical basis \cite{gottesman1997stabilizer} in which $\bar{X}_i$ and $\bar{Z}_j$ intersect on exactly $\delta_{ij}\in\{0,1\}$ qubits. (b) Our goal is to measure a set of logical Paulis that, when decomposed in the standard basis, are qubit-wise-commuting, by which we mean that the operators' supports on each logical qubit are either equal or identity. (c,d) Use LDPC surgery techniques, brute-force branching~\cite{ZL} and thickening~\cite{williamson2024low}, to create completely disjoint representatives of the logical basis operators that are required. (e) Apply gauging measurement \cite{williamson2024low}, possibly with adapters \cite{swaroop2024universal} between disjoint components, to measure the desired set of logical Paulis. Most generally, to instead measure arbitrary commuting sets of logical Paulis, we can use this core construction a constant number of times in sequence, see Section~\ref{sec:commuting_products}.}
    \label{fig:summary}
\end{figure}

At a high level, the overall space overhead of our scheme comes from the fact that brute-force branching $t$ logicals, with weight up to $\omega$ each, costs $\CO(t\omega \log t)$, and then measurement requires a further $\CO(t\omega\log^3\omega)$. The overall scheme described in this paper is summarised in Fig.~\ref{fig:summary}.

There are two ways to extend our scheme to measure more general collections of Pauli operators.
\begin{enumerate}
    \item While we state our result in terms of logically disjoint products, the same procedure applies to non-disjoint Pauli products that have the same or identity action on every logical operator. For example, we can measure $\bar{X}_1\bar{Z}_2\bar{I}_3$ simultaneously with $\bar{I}_1\bar{Z}_2\bar{X}_3$, but not $\bar{X}_1\bar{Z}_2\bar{I}_3$ with $\bar{Z}_1\bar{X}_2\bar{I}_3$. In this slightly more general case, the asymptotic space overhead of $\CO\big(T\omega(\log T+ \log^3\omega)\big)$ is the same except $T$ is now the total number of non-identity logical Pauli terms in the Pauli products (e.g.~$T=4$ for the above example, the sum of the logical weights of $\bar{X}_1\bar{Z}_2\bar{I}_3$ and $\bar{I}_1\bar{Z}_2\bar{X}_3$).
    \item In Section~\ref{sec:commuting_products}, we extend our scheme to measure arbitrary commuting subgroups of logical Pauli operators, using twist-free surgery techniques from Refs.~\cite{CC} and~\cite{ZL}. 
    These techniques require ancilla logical states which can be encoded in the same code block or additional code blocks.
    The ability of our scheme to directly measure logical $Y$ terms in parallel significantly reduces the overhead in preparation of such logical ancillary states as compared to Ref.~\cite{ZL}.
\end{enumerate}

\begin{figure}
    \centering
\begin{tabular}{ |c||c|c|c|c|c|  }
 \hline
 Scheme & Refs.~\cite{Coh, cross2024improved}, & Ref.~\cite{ZL} & Refs.~\cite{williamson2024low, swaroop2024universal, ide2024fault} & This work \\
 
 Space overhead & $\CO(t\omega d)$  &  $\tilde{\CO}(t\omega d)$ & $\tilde{\CO}(t\omega)$ & $\tilde{\CO}(t\omega)$\\

  Parallel & \xmark & \checkmark & \xmark & \checkmark\\

 $S$ gate-free & \checkmark &\xmark &\checkmark & \checkmark\\

Stabiliser & \xmark & \xmark & \checkmark & \checkmark \\

Logical ancilla-free &\checkmark &\xmark &\checkmark & \checkmark\\

 \hline
\end{tabular}

    \caption{Comparison of different surgery schemes. The \textit{space cost} is the asymptotic space overhead of a scheme to measure $t$ logical Pauli operators with maximum representative weight $\omega$ while maintaining the code distance $d$. Logarithmic factors are suppressed by $\tilde{\CO}(\cdot)$ notation. A scheme is labelled \textit{parallel} when, for a generic LDPC code and set of logical operators to be measured, the scheme maintains the LDPC property while measuring the logical operators in $\CO(1)$ logical timesteps, each taking up to $\CO(d)$ rounds of syndrome extraction. A scheme is labelled \textit{$S$ gate-free} when it does not require transversal $S$ gates or $\ket{Y}$-state distillation. A scheme is \textit{stabiliser} when it is applicable to generic stabiliser codes, rather than just CSS codes. A scheme is \textit{logical ancilla-free} when the scheme can be applied without requiring any \textit{logical} ancilla qubits on top of the prior \textit{physical} qubit space overhead.}
    \label{fig:comparison_table}
\end{figure}

In Fig.~\ref{fig:comparison_table}, we compare prior schemes to our own. Column 1 is the CKBB scheme~\cite{Coh} and similar schemes, including Ref.~\cite{cross2024improved}. Column 2 contains the parallel measurement schemes from Ref.~\cite{ZL}, which use brute-force branching and devised sticking respectively. As brute-force branching has a lower asymptotic overhead we use that overhead for the comparison. Column 3 contains the various schemes based on expander graphs, Refs.~\cite{williamson2024low, swaroop2024universal, ide2024fault}. There are many subtleties not included in the table. For example, in columns 1 \& 2, Refs.~\cite{Coh, ZL} require moving to a subsystem code, which can prevent a threshold, while Ref.~\cite{cross2024improved} is not guaranteed to be LDPC, but can obtain improved overheads in practical cases. In column 3, Refs.~\cite{williamson2024low, swaroop2024universal} apply to generic stabiliser codes, while Ref.~\cite{ide2024fault} is specified only for CSS codes. Lastly, there are two different definitions of parallelism used in Ref.~\cite{ZL}:
\begin{enumerate}
    \item ``Conventional" parallelism, whereby logical operations are applied simultaneously if they act on different logical qubits, see Def.~\ref{def:logical_disjoint}.
    \item ``Ultimate" parallelism, whereby logical operations are applied simultaneously if they commute.
\end{enumerate}
In particular, in the latter case the logical operations may overlap on logical qubits. While the majority of this work focusses on the former case, which is reflected in Figs.~\ref{fig:parallel_circuit} and
\ref{fig:comparison_table}, we also extend our results to the latter in Sec.~\ref{sec:commuting_products}. In that extension, the scheme is no longer logical ancilla-free, but all the remaining rows in Fig.~\ref{fig:comparison_table} are identical.

\begin{remark}[Weights of logical representatives]
Existing methods of LDPC code surgery deform the local code around logical representatives, hence the cost in terms of additional qubits and checks is reliant on the choice of representatives and their weights.

While there are codes for which such Pauli representatives can always be chosen to have weight $d$, or weights bounded by $\CO(d)$ as the blocklength increases, this is not true in general. 3D toric codes, for example, have $\bar{X}$ logical operators which are asymptotically larger than $d$ \cite{HZW}.

As a consequence, to be fully general, our costs must be computed in terms of $\omega$, the largest logical representative weight of the Paulis to be measured, rather than $d$.
Furthermore, 
our surgery scheme requires the logical operator representatives $\bar{X}_1, \cdots, \bar{X}_k$ and $\bar{Z}_1, \cdots, \bar{Z}_k$ to satisfy a set of conditions regarding their supports. 
While such a basis always exists in any stabiliser code \cite{gottesman1997stabilizer}, these representatives could have weights that are asymptotically far larger than $d$, bounded by $\CO(n)$ in general. This is unsurprising, as we require no additional structure of the LDPC code which we are performing code surgery on, and good quantum LDPC codes are known to exist~\cite{PK}, hence the logical representative weights to be measured cannot be bounded in general by anything less than $\CO(n)$.
\end{remark}

\subsection{The setting}\label{sec:setting}
First, we assume that we can modify the qubit connectivity between rounds of logical measurements, so we can initialise new ancilla systems with very different connectivity for each round of logical measurements without concern. 
This is different from the setting of Refs.~\cite{CHRY, he2025extractors, yoder2025tour}, in which ancilla systems with fixed connectivity are constructed that can be toggled to measure different serialised Pauli products.

Second, we focus on the asymptotic regime, rather than considering finite-size codes with specific (pseudo-)thresholds. 
For finite-size codes, we expect that our procedure can be substantially optimised to reduce the overhead far below what the asymptotic upper bound indicates, following Refs.~\cite{Cow24, CHRY}. 
For example, it has been found that a variety of near-term codes do not require decongestion of the expander graphs used for measurement~\cite{IGND, WY}. Furthermore, while brute-force branching is asymptotically efficient, the constant factor overheads are formidable and so minimising the branching cost is vital for our scheme to be efficient on low-blocklength codes.
We leave such numerical optimisation to future work.

Third, the noise model we consider in our notion of fault tolerance is a phenomenological Pauli and measurement noise model, where Pauli faults may occur on data qubits or measurement outcomes. The phenomenological noise model is justified in Appendix~\ref{app:ft_proofs}.

\section{Preliminaries}\label{sec:prelims}

\subsection{Stabiliser codes and Tanner graphs}
Here we introduce some basic definitions and lemmas, which are used later.

\begin{definition}
Let $\CP^n$ be the Pauli group over $n$ qubits. A qubit stabiliser code $Q$ is specified by an Abelian subgroup $\CS \subset \CP^n$ such that the codespace $\CH$ is the mutual $+1$ eigenspace of $\CS$, i.e.
\[U \ket{\psi} = \ket{\psi}\quad \forall\ U \in \CS,\ \ket{\psi}\in \CH.\]
We say $\CS$ is the stabiliser group for $Q$.
\end{definition}

A qubit stabiliser code $Q$ can be specified by a stabiliser check matrix 
\[H =[H_X | H_Z ] \in \F_2^{r\times 2n},\ \textrm{such that}\ H\begin{pmatrix}0 & 1 \\ 1 & 0\end{pmatrix}H^\intercal = 0,\]
where a row $[u | v]$ corresponds to a generator of the stabiliser group, and therefore check on $Q$, $i^{u\cdot v}X(u)Z(v)$, for $u, v \in \F_2^n$.

\begin{definition}
A qubit CSS code $Q$ is a qubit stabiliser code where the generators of $\CS$ can be split into two sets $\CS_X$ and $\CS_Z$. $\CS_X$ contains Pauli products with terms drawn from $\{X,I\}$ and $\CS_Z$ terms drawn from $\{Z,I\}$.
\end{definition}

Thus there is a stabiliser check matrix $H$ for $Q$ such that
\[H = \begin{pmatrix}H_X & 0 \\ 0 & H_Z\end{pmatrix},\qquad H_X H_Z^\intercal = 0.\]

We commonly conflate a Pauli $Z(v)$ operator with its vector $v$ in $\F_2^n$, and the same for a Pauli $X(u)$ operator.

\begin{definition}
A low-density parity check (LDPC) code family is an infinite family of
stabiliser codes such that the number of qubits participating in
each check operator and the number of stabiliser checks that
each qubit participates in are both upper-bounded by a constant.
\label{def:ldpc}
\end{definition}

Logical operators in a stabiliser code are Pauli products that commute with all the checks of the code.
As the stabiliser group is Abelian, all stabilisers are equivalent to the trivial logical operator.
The code distance $d$ of $Q$ is the minimum Pauli weight of all the nontrivial logical operator representatives.

\begin{figure}[t]\centering
\[\input{tikz_files/shor_tanner.tikz}\]
\caption{A Tanner graph for the $\llbracket 9,1,3 \rrbracket$ Shor code.}
\label{fig:ShorTanner}
\end{figure}

We can use Tanner graphs to describe stabiliser codes, see Figure~\ref{fig:ShorTanner}.

\begin{definition}\label{def:tanner_graph}
A Tanner graph is a bipartite graph with a vertex for each qubit and each check. A qubit is connected to each check in which it participates
with an edge labeled $[1|0]$, $[0|1]$, or $[1|1]$ depending on whether the check acts on the qubit as $X$, $Z$, or $Y$.
\end{definition}

In our convention, data qubits are black circles and checks are boxes. 
If a check only measures $Z$ or $X$ terms then we omit the edge label, and instead label the box with a $Z$ or $X$. 
Therefore for a CSS code we have no edge labels, and all boxes contain either $Z$ or $X$ labels. 
In this case, the condition that all stabilisers commute is the same as saying that the number of paths between every $Z$ and $X$ check is even.

Depicting large Tanner graphs directly becomes unwieldy, so we use \textit{scalable} notation\footnote{Terminology borrowed from Ref.~\cite{CHP}.} to make them more compact. Qubits are gathered into disjoint named sets $\CQ_0, \CQ_1, \CQ_2, ...,$ and the same for checks $\CC_0, \CC_1, \CC_2, ...\,$. An edge between $\CQ_i$ and $\CC_j$ is then labeled with the stabiliser check matrix of $Q$ restricted to $\CQ_i$ and lifted to $\CC_j$, so we have
\[ [C_X|C_Z] \in \F_2^{|\CC_j|\times 2|\CQ_i|}.\]
If that check matrix is all zeros then omit the edge entirely. If the checks are all of the same type, then we omit the all-zeros half of the check matrix; for example if $\CC_j$ contains only $Z$ checks then we label the check box $Z$ and the edge
\[ [C_Z] \in \F_2^{|\CC_j| \times |\CQ_i|}.\]

\begin{figure}[tb]
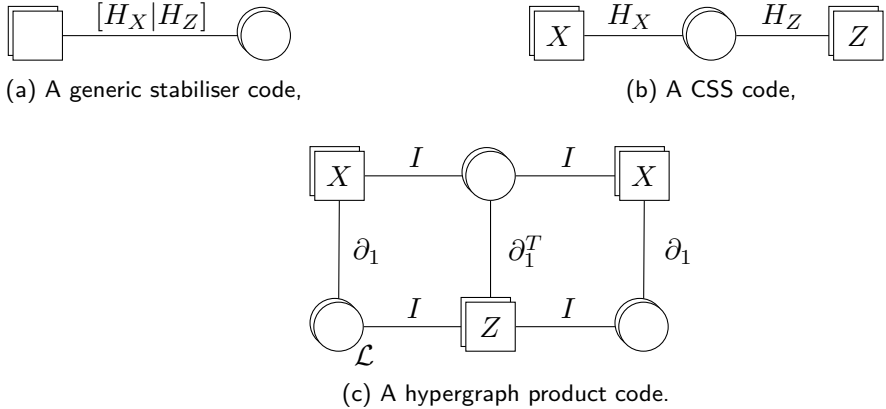

    \hfill
     \begin{subfigure}[t]
     {0.36\textwidth}
         \centering
         \input{tikz_files/tanner_eg1.tikz}
         \caption{A generic stabiliser code,}
     \end{subfigure}
     \begin{subfigure}[t]{0.6\textwidth}
     \centering
         \input{tikz_files/tanner_eg2.tikz}
             \caption{A CSS code,}
     \end{subfigure}
     \hfill
     \vspace{5mm}
     \begin{subfigure}[t]{\textwidth}
     \centering
     \input{tikz_files/hypergraph_prod.tikz}
     \caption{A hypergraph product code.}
     \end{subfigure}
     \caption{Examples of scalable Tanner graphs.}
         \label{fig:example_tanner_graphs}
\end{figure}

For example, in Figure~\ref{fig:example_tanner_graphs} we show some basic scalable Tanner graphs with qubit sets drawn as circles and check sets drawn as boxes.
We have (a) a generic stabiliser code, (b) a CSS code, and (c) the hypergraph product~\cite{TZ} of a classical code $C$, with bits $\mathcal{L}$ and check matrix $\del_1$, by a repetition code $R$ with blocksize 2 and check matrix $\begin{pmatrix}
    1 & 1
\end{pmatrix}$. 
See Appendix~\ref{app:hypergraph_prods} for a concise summary of hypergraph product codes.
We use Tanner graphs, particularly scalable Tanner graphs, extensively to describe our procedures.

Throughout this work, we build on prior literature on quantum LDPC code surgery. 
We now recount relevant elements that we use.

\subsection{Brute-force branching}\label{sec:branching}
\begin{figure}\centering
\includegraphics[scale=0.16]
{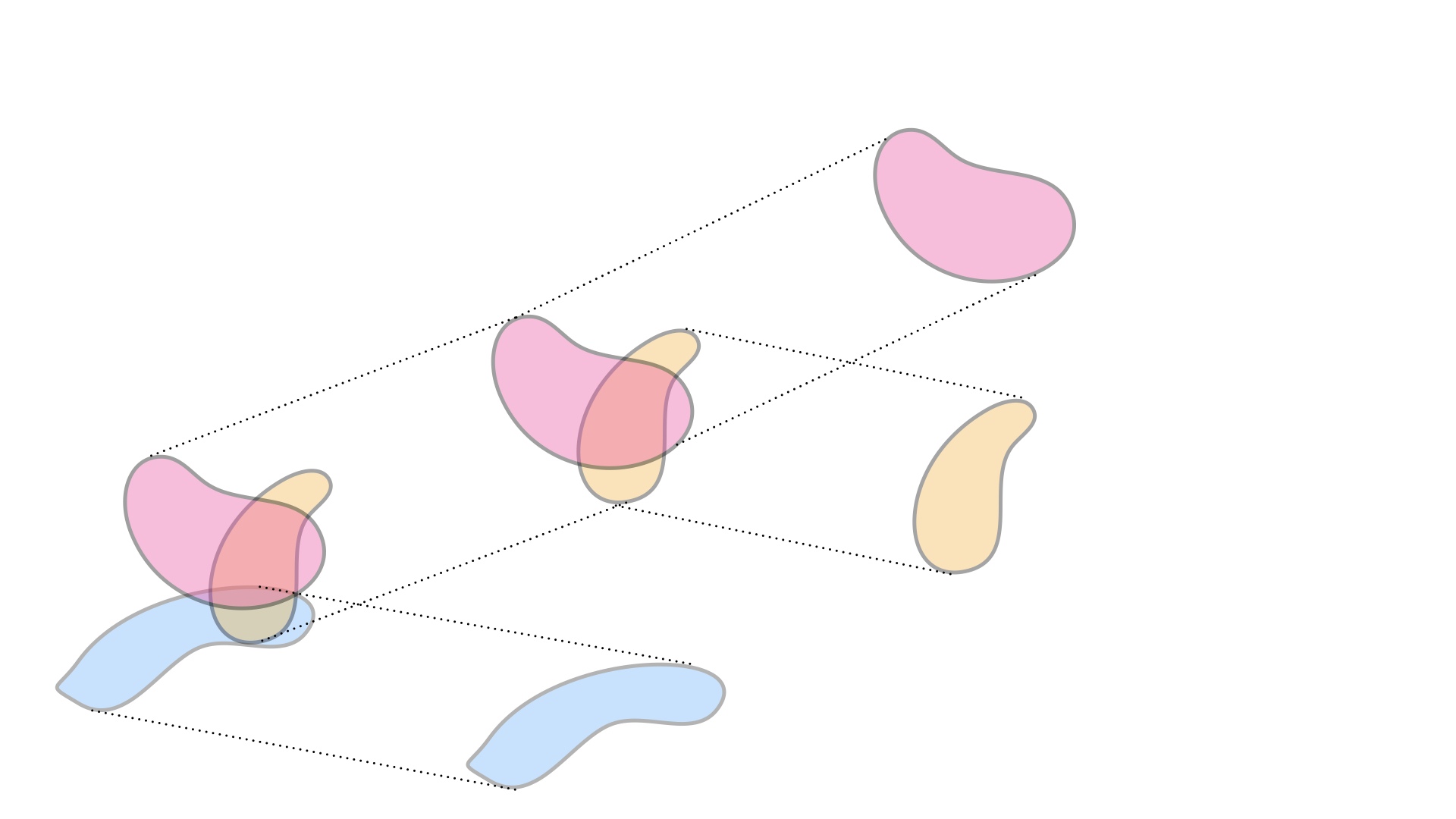}
\caption{Illustration of brute-force branching logical representatives with overlapping support.}\label{fig:logical_branching}
\end{figure}

Brute-force branching is a method for separating logical operators into new, disjoint supports on the leaves of a tree, see Figure~\ref{fig:logical_branching}. These new operators are equivalent to the original, but are more convenient for performing logical measurements~\cite{ZL} without losing the LDPC property. After logical measurements, the branches are removed by measuring them out. Branching does not perform any logical operation on the code.

Brute-force branching is formally defined as follows: let $Q$ be a CSS code. Let $\mathcal{I} = \{v_i : i \in 1,2,...,t\}$ be a set of $\bar{Z}$ logicals with support on some data qubits, and let $V = \bigcup_i v_i$ be the union of those supports. Any logical can overlap with any other, and there may be other logicals $\mathcal{J}$ with support wholly in $V$ other than the $v_i$ logicals.

A \textit{brute-force branching} of $V$ is then a tree, which we always take to be a binary tree, of small hypergraph product patches, called branching stickers, connected to $V$ in such a manner as to separate all the logicals $v_i$ into disjoint supports. $V$ is then the root of the tree, and each leaf contains a copy of each $v_i$.

\begin{figure}
\[\input{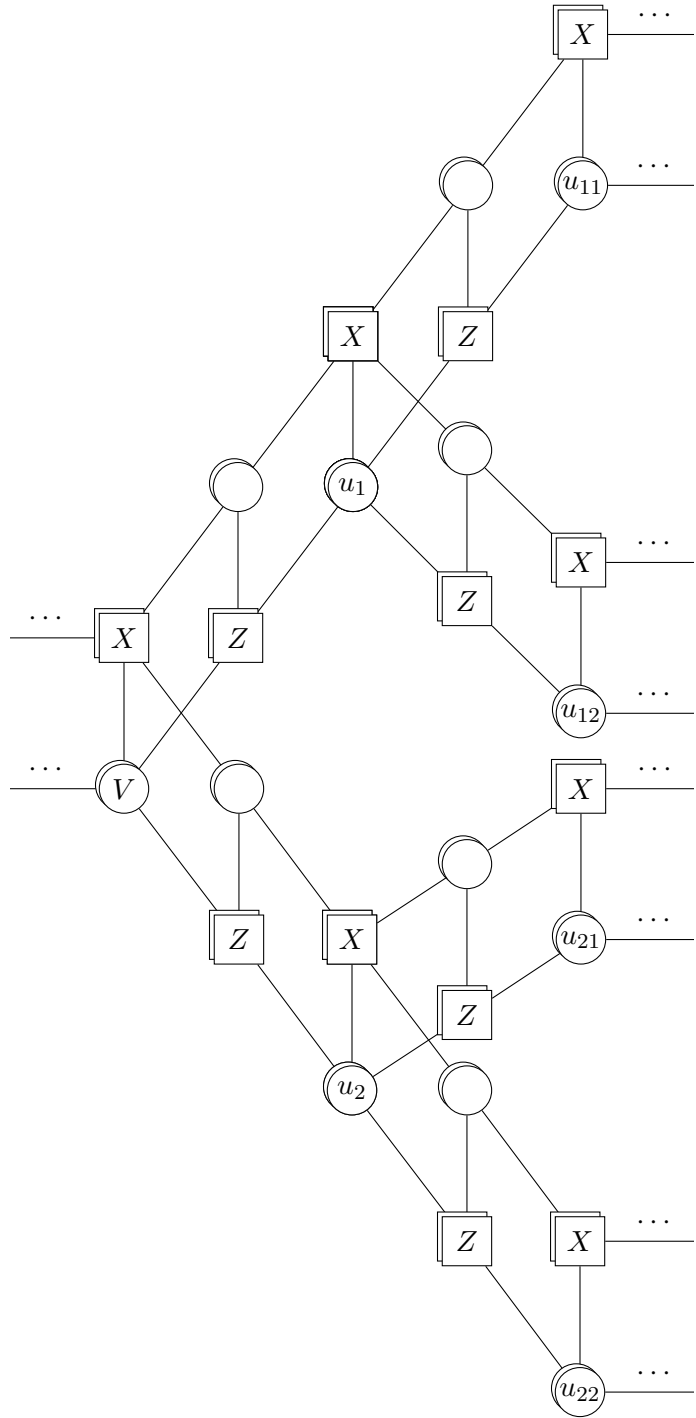}\]
\caption{Scalable Tanner graph for brute-force branching. Starting with a union of logicals $V$, each layer of the binary tree splits a union of logicals in two, until we are left with disjoint logicals on leaves. In this Tanner graph we do not include the checks and data qubits in the rest of the initial code $Q$, leaving only dangling edges on the l.h.s. to suggest their existence. Similarly, we leave dangling edges on the r.h.s. to suggest the existence of further branching layers.}\label{fig:all_branching}
\end{figure}

Brute-force branching can be described by a scalable Tanner graph, as shown in Figure~\ref{fig:all_branching}. Explicitly, let $\mathcal{T}(Q)$ be the Tanner graph of $Q$. Let $|\CI| = t$. Take $\lceil \frac{t}{2} \rceil$ of the logicals in $\CI$ and let the union of their supports on data qubits be $S_1$. Let the set of all $X$-checks incident to $S_1$ be $\chi_1$. Including the edges $E_1$ between $S_1$ and $\chi_1$, these generate a bipartite subgraph $G_1(S_1\cup \chi_1, E_1)$ of the Tanner graph of $Q$. 
Then introduce a new graph $G_1^\intercal(S_1^\intercal \cup \chi_1^\intercal, E_1)$. This graph has the same vertices and edges as $G_1$, but each vertex in $S_1^\intercal$ is a $Z$-check, rather than a data qubit, and each vertex in $\chi_1^\intercal$ is a data qubit, rather than an $X$-check. Connect $G_1^\intercal(S_1^\intercal \cup \chi_1^\intercal, E_1)$ to the original Tanner graph of $Q$ by the introduction of edges $(u, u')$ for $u \in S_1$ and its corresponding $u' \in S_1^\intercal$ for each $u$, and the same for $v \in \chi_1$ and $v' \in \chi_1^\intercal$.
Now introduce another new graph $H_1$, which is a copy of $G_1$. Connect $H_1$ to $G_1^\intercal$ via edges $(u', u)$ for each $u' \in S_1^\intercal$ and $(v', v)$ for each $v \in \chi_1^\intercal$. Thus far, we have created a single branch for $\lceil \frac{t}{2} \rceil$ of the logicals in $\CI$. The creation of this single branch can be viewed as gluing a hypergraph product code, see Fig.~\ref{fig:example_tanner_graphs}(c), to the code $Q$.

Repeat this procedure for the other $\lfloor \frac{t}{2} \rfloor$ logicals: given the union of their supports on data qubits $S_2$ and $\chi_2$, all $X$-checks incident to $S_2$, construct $G_2^\intercal(S_2^\intercal \cup \chi_2^\intercal, E_2)$, connect it in a 1-to-1 fashion, then construct $H_2$ and connect it to $G_2^\intercal$ in a 1-to-1 fashion.\footnote{Throughout this paper the branching ancillary system has 1-to-1 connections from the first layer of a branch to the appropriate subset of logicals being branched. In Ref.~\cite{ZL} this is left as a variable, so one can choose this connectivity.}

Repeat this process $\CO(\log t)$ times until we have our $t$ logicals on different stickers on the leaves of the tree. Each leaf corresponds to an equivalent copy of a logical $v_i \in \mathcal{I}$. Hence all $t$ logicals have disjoint support, and we have preserved the LDPC property. If the initial logicals had maximum representative weight $\omega$ then each level of the tree uses $\CO(t\omega)$ qubits and checks, so the number of additional qubits and checks required for branching will be at most $\CO(t\omega\log t)$.

Of course, if there are logicals in $V$ which have no overlap with any other logicals $v_i$ , they do not need to be brute-force branched and can be measured directly. If the initial code $Q$ has a basis of logical operators of both $\bar{Z}$ and $\bar{X}$ type which are disjoint where they commute, as with e.g. disjoint surface code patches, then brute-force branching is unnecessary.

In Appendix~\ref{app:example_branching} we demonstrate branching in detail on toy examples.

\begin{lemma}\label{lem:branch_no_new}
Brute-force branching creates a new representative on each leaf for each $v_i\in \mathcal{I}$, and introduces no new logical operators.
\end{lemma}
\proof
Here we give a proof sketch for readers familiar with the CKBB scheme~\cite{Coh}. In Appendix~\ref{app:branch_cylinders} we introduce a new perspective on brute-force branching as a series of mapping cylinders. This provides an algebraic and formal proof, but which requires some knowledge of elementary homological algebra.

For each $v_i \in \mathcal{I}$, consider the path $P$ down the binary tree from the root to its corresponding leaf. Multiply $v_i$ by the $Z$ checks connected to data qubits in $v_i$ on the first layer, on the branch in the path $P$. As $v_i$ is a logical operator, this product of $Z$ checks leaves no support in layer 1, and cleans $v_i$ to one of the branches in layer 2. Continue this procedure along $P$, until all support has been moved to the leaf corresponding to $v_i$. This logical is then another representative of the same operator as $v_i$.

When measuring $\bar{Z}$ logicals in the CKBB scheme, operators corresponding to new logical qubits (gauge logicals) appear precisely at every odd layer of the attached Tanner graph, that is $Z$ layers~\cite{Coh}. Let $\del_1$ be the local code of $\bar{Z}$ logicals and incident $X$ checks. Then, in the CKBB scheme, the data qubits on odd layers and their incident $Z$ checks form a code $\del_1^\intercal$. As $\ker(\del_1^\intercal) \neq 0$ in general, these can form new gauge $\bar{X}$ logicals, and indeed these are the only new logicals that can appear, other than their anticommuting $\bar{Z}$ partner logicals~\cite{Coh}.

When performing brute-force branching, all these new gauge logicals are actually stabilisers. For any new gauge logical $g$ in an odd layer $l$, apply the corresponding $X$ checks in layer $l+1$. This creates no new $X$ support on layer $l+1$, as $g \in \ker(\del_1^\intercal)$ by definition, but does create some $X$ support on layer $l+2$. This new $X$ operator must also not violate any $Z$ checks on this layer, as it is a product of a logical and a stabiliser. Continue applying $X$ checks until the last layer, at which point there are no new layers upon which to add $X$ support, and hence the gauge logical was a product of $X$ checks, and so a stabiliser.

This proof is essentially a cleaning argument, cf. Ref.~\cite[Lem.~1]{BT}.
\endproof

\begin{remark}
As a consequence of Lemma~\ref{lem:branch_no_new}, unlike in the CKBB measurement scheme, we do not have to move to a subsystem code when brute-force branching. The CKBB measurement scheme for a $\bar{Z}$ operator ends with a layer of $Z$-checks, rather than a layer of $X$-checks when brute-force branching, which leads to gauge operators which are inequivalent to stabilisers.
\end{remark}

\begin{lemma}\cite[Thm.~2]{ZL}\label{lem:branching_distance}
Brute-force branching cannot reduce the code distance.
\end{lemma}
\proof
Assume we are branching $\bar{Z}$ logicals w.l.o.g. on a starting code $Q$ with distance $d$. As there are no new logical qubits, to prove that the code distance is preserved we only need to show that multiplying old logicals in the starting code by stabilisers in the new code does not reduce the distance.

For the $X$ distance, observe that all new $X$ checks are on new data qubits, but there are old $X$ checks which have been deformed to have support on the first layer of the branching. Applying these old $X$ checks cannot reduce the weight of an initial $\bar{X}$ operator's support on $Q$ below $d$, as it only has new support on new qubits. In other words, if we take an operator $\bar{\Lambda}_X$ on $Q$ and apply deformed checks $C_D$, the restriction of $C_D\bar{\Lambda}_X$ to $Q$ must have weight at least $d$. The new $X$ checks only have support on new qubits, and so cannot reduce the weight of any operator on $Q$. Therefore the $X$ distance is at least $d$.

For the $Z$ distance, we start with an arbitrary operator $\bar{\Lambda}_Z$ on $Q$. Applying a new $Z$ check in the first layer moves the support of $\bar{\Lambda}_Z$ from support of the check in $Q$ to their matching data qubits in the second layer of that branch, which is an $X$ layer, potentially leaving behind more support in $Q$ and on the first layer. We can repeat this process, but there are always $X$ layers after every $Z$ layer so we always add more support, and so the weight of $\bar{\Lambda}_Z$ cannot be decreased, hence the $Z$ distance is at least $d$.
\endproof

\begin{remark}
The reason why the CKBB scheme~\cite{Coh} requires $\CO(d)$ layers to preserve the distance, rather than only 2 layers for each branch when brute-force branching, is also because in the CKBB measurement scheme for a $\bar{Z}$ operator, we end with a $Z$ layer rather than an $X$ layer.
\end{remark}

To perform brute-force branching of $\bar{Z}$ logicals, we initialise all new qubits in the $\ket{+}$ basis at the same time and then measure all checks. For $\bar{X}$ logicals, the scheme is dualised and we initialise in the $\ket{0}$ basis instead.

\begin{lemma}\label{lem:branch_fault_tolerance}
Brute-force branching is a fault-tolerant operation with fault-distance $d$, assuming we use at least $d$ rounds of stabiliser measurements. Un-branching by measuring out ancilla qubits and measuring all stabilisers in the original code is also a fault-tolerant operation, but only requires one round of stabiliser measurements.
\end{lemma}
\proof
See Appendix~\ref{app:branching_ft}.
\endproof

Despite brute-force branching originally being defined for $X$ or $Z$-type operators on a CSS code, it is possible on any stabiliser code under a suitable condition given below. This observation is similar to the approach in Ref.~\cite{WY} for measurement, but for branching instead.

\begin{lemma}\label{lem:mixed_branching}
    Let $Q$ be a stabiliser code. Let $\CI = \{\bar{\Lambda}_1,\bar{\Lambda}_2,\cdots,\bar{\Lambda}_t\}$ be a set of $t$ logical Pauli representatives such that for each $i,j\in [t]$ and data qubit $\alpha$ in $Q$ either (a) $\bar{\Lambda}_i$ and $\bar{\Lambda}_j$ act on $\alpha$ with the same Pauli operator or (b) at least one of $\bar{\Lambda}_i$ and $\bar{\Lambda}_j$ act with identity on $\alpha$.
    Then all $t$ logical representatives can be simultaneously branched onto leaves.
\end{lemma}
\proof
For each data qubit $\alpha$ in $V = \bigcup_i \bar{\Lambda}_i$, apply a suitable single-qubit Clifford $C_\alpha$ to convert the logical representative into a $Z$ Pauli on that qubit. This permutes the bases of incident checks accordingly. As the logical representatives act with the same Pauli, or the identity, this simultaneously changes the basis of every logical representative in $\CI$ on data qubit $\alpha$ to $Z$ or $I$.

Now, every check on the code which locally anticommutes with the action of any $\bar{\Lambda}_i$ on a given qubit $\alpha$ has a $Y_\alpha$ or $X_\alpha$ term. Considering only the $X_\alpha$ component of a $Y_\alpha$ term, construct the branch accordingly as for $\bar{Z}$ logicals in a CSS code. Last, convert each data qubit back to the original basis by applying $C_\alpha^\dagger$.

As the checks all commute in the $Z$ basis, they must also commute when permuted back to their original basis. Because there exists a product of checks which cleans each logical $\bar{\Lambda}_i$ to a leaf in the $Z$ basis, the same product of checks must clean that logical to a leaf in the original basis.
\endproof

The basis changing is a step performed offline to compute the checks and connectivity required for the branches; the local Cliffords are not intended to be physically applied on a quantum computer.

We illustrate the branch for $S_1 \subset V$ once the basis has been changed to the $Z$ basis below.
\[\input{tikz_files/all_branching_stab.tikz}\]
$[A_X | A_Z]$ is the check matrix of the code $Q$ restricted to those checks with $X$-support in $S_1$. Dangling edges are left on the l.h.s. to indicate connections to the rest of the code, and on the r.h.s. to indicate connections to the rest of the branch. There is a similar, overlapping branch for $S_2 \subset V$. Observe that this $S_1$ branch is identical to $\bar{Z}$ logical branching in a CSS code, albeit some incident checks to $S_1$ in the original code have some $Z$ check component, because they involve measurements in the $Y$ basis. This $Z$ check component on the original code does not affect the commutation relations or multiplication of logical operators through to leaves by $Z$ checks in odd layers.

Because this more general branching is equivalent to branching $\bar{Z}$ logicals up to local Cliffords and an irrelevant $A_Z$ check component on the original code, all previous lemmas from this section apply.
Another consequence of this generalised branching is that each logical representative on a leaf will be a product of $Z$ Paulis, regardless of the basis it acted on in the original code -- whether it is a $\bar{Z}$, $\bar{X}$ or $\bar{Y}$ operator.

\subsection{Gauging logical measurements}\label{sec:gauging}
Our procedure for fault-tolerant quantum computation also involves gauging logical measurements~\cite{WY}. Gauging logical measurements can be applied to any stabiliser code. In the case where a code is CSS the gauging logical measurement framework coincides with the homological measurement framework of Ref.~\cite{IGND}.

\begin{definition}\cite{WY}\label{def:gauging_measurement}
Let $\CL$ be a $\bar{Z}$ logical operator in a stabiliser code. A measurement of $\CL$ in the \textit{gauging logical measurement} framework uses a set of appended ancillary qubits and checks forming a graph $\CG(\CV,\CE,\CF)$. 

$\CG(\CV,\CE,\CF)$ has a set of vertices $\CV$, which are identified with $Z$ checks, in 1-to-1 correspondence with data qubits of $\CL$. Each vertex has support on that data qubit in $\CL$ given by the correspondence. $\CG$ also has a set of edges $\CE$, which are identified with data qubits, and some of these are included in deformed $X$ checks which are incident to $\CL$. Lastly, $\CG$ has a set of faces $\CF$ which form a cycle-basis of $\CG$. These faces are for gauge-fixing, are only connected to edges in $\CE$, and do not interact with the original code.

Using a scalable Tanner graph, we have:
\[\input{tikz_files/gauging_measurement.tikz}\]
where $\CL$ is the original logical operator, and $\CS$ is the set of all $X$-checks incident to $\CL$. The sets $\CV$, $\CE$ and $\CF$ are newly introduced to perform the measurement, and
\begin{itemize}
    \item $F$ is a matrix defining the connectivity from $Z$ checks to data qubits in $\CL$; for our purposes $F = I$.
    \item $G$ is the incidence matrix of the graph $\CG$.
    \item $M$ defines the connectivity from $X$ checks incident to $\CL$ to new data qubits. $M$ is determined by the choice of graph and initial logical operator.
    \item $N$ picks out a cycle basis of $\CG$, such that $\ker{G} = \im N^\intercal$.
\end{itemize}

To perform the measurement, initialise all new edges in $\CG$ in the $\ket{+}$ basis, and measure all stabilisers of the deformed code.
\end{definition}

The auxiliary graph $\CG$ must satisfy some desiderata to yield a favorable code surgery scheme.

\begin{theorem}\cite{WY, SJY}\label{thm:desiderata}
To ensure the deformed code has exactly one less logical qubit than the original code and measures the target logical operator $\bar{Z} = Z(\CL)$, it is sufficient that
\begin{enumerate}
\setcounter{enumi}{-1}
    \item $\CG$ is connected.
\end{enumerate}
To ensure the deformed code is LDPC, it is necessary and sufficient that
\begin{enumerate}
\setcounter{enumi}{0}
    \item $\CG$ has $\CO(1)$ vertex degree.
    \item Each stabiliser of the original code has a short perfect matching in $\CG$, and each edge is in $\CO(1)$ matchings.
    \item There is a cycle basis of $\CG$ in which (a) each cycle is length $\CO(1)$ and (b) each edge is in $\CO(1)$ cycles.
\end{enumerate}
To ensure the deformed code has code distance at least the distance $d$ of the original code, it is sufficient that
\begin{enumerate}
\setcounter{enumi}{3}
    \item $\CG$ has relative Cheeger constant $\beta_d(\CG, f(\CL)) \geq 1$.
\end{enumerate}
\end{theorem}

We elide the details of the expansion properties and perfect matchings, as they are not the focus of this work. It is sufficient for our purposes to say that, given any logical operator, these desiderata can always be satisfied, and in fact Ref.~\cite{williamson2024low} proves that it is easy to do so.

The space overhead of such a graph for measuring any arbitrary logical operator with weight $\CO(\omega)$ is $\CO(\omega\log^3\omega)$. The $\log^3\omega$ factor comes from decongestion~\cite{FH}, whereby the graph is thickened in order to find a low-weight cycle-basis, a procedure which can be performed efficiently by a randomised algorithm. By thickening, we add ``dummy vertices'' which are not connected to the original code $Q$ but are required to maintain the LDPC property.

\begin{remark}
    The desiderata above ensure that the deformed code is LDPC when measuring only one target logical operator, or $t$ disjoint logical operators. If there are many representatives to be measured, and they overlap substantially, then Theorem~\ref{thm:desiderata} is not sufficient to preserve the LDPC property, as data qubits at the overlap are involved in $\CO(t)$ incident checks.
\end{remark}

\begin{lemma}\label{lem:gauging_fault_tolerance}\cite[Lem.~7]{WY} If all the desiderata of Theorem~\ref{thm:desiderata} are satisfied, the gauging measurement procedure given by initialising all new data qubits in suitable bases and measuring all stabilisers for at least $d$ rounds of measurement has fault-distance $d$.
\end{lemma}
\subsection{Universal adapters}

The last definition we need is that of adapters, whereby ancilla systems for measuring disjoint logical Paulis using the gauging logical measurements formalism can be connected to measure their product. 

\begin{definition}\cite{SJY}
Consider two disjoint logical operators with accompanying graphs $\CG_l$ and $\CG_r$ defining gauging logical measurement systems for each individual operator. Then a universal adapter $\CA$ is a set of additional vertices, edges and faces connected to $\CG_l$ and $\CG_r$. This converts $\CG_l$ and $\CG_r$ into a larger, \textit{adapted} graph $\CG_l \sim_{\CA} \CG_r$.
\end{definition}

\begin{figure}[t]
    \centering
    \input{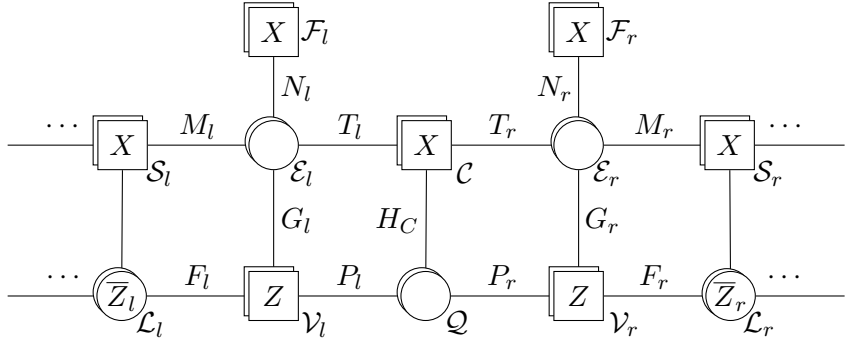}
    \caption{Measuring $\bar{Z}_l\bar{Z}_r$ without measuring either logical operator individually. The adapter edges joining auxiliary graphs $G_l$ and $G_r$ into one larger auxiliary graph are those hosting the qubits $\mathcal{Q}$. The matrices $T_l,T_r,P_l,P_r$ are efficiently determined using the $\mathsf{SkipTree}$ algorithm,~\cite[Thm.~7]{SJY}. We do not include in this diagram qubits or checks outside the supports of the logicals $\bar{Z}_l$ and $\bar{Z}_r$, and instead suggest their existence through dangling edges.}
    \label{fig:joint_measurement}
\end{figure}

\begin{theorem}\cite[Thm.~11]{SJY} \label{thm:adapters}
Provided auxiliary graphs to perform gauging measurements of non-overlapping logical operators $\bar{Z}_0,\bar{Z}_1,..., \bar{Z}_{t-1}$ that satisfy the graph desiderata of Theorem~\ref{thm:desiderata}, there exists an auxiliary
graph to perform gauging measurement of the product $\bar{Z}_0\otimes \bar{Z}_1\otimes ...\otimes \bar{Z}_{t-1}$ satisfying the desiderata of Theorem~\ref{thm:desiderata}.\footnote{The first version of this theorem required the logical operators to be irreducible but was later generalised to remove this requirement.}
\end{theorem}

This gauging measurement first takes the set of operators $\bar{Z}_0,\bar{Z}_1,..., \bar{Z}_{t-1}$ and applies adapters as in Figure~\ref{fig:joint_measurement} between $t-1$ overlapping pairs of disjoint logicals, for example $\bar{Z}_0\bar{Z}_1$ then $\bar{Z}_1\bar{Z}_2$ etc., until all $t$ logicals are connected. In particular, the measurement can be performed while maintaining the LDPC property and fault-distance of the code.

This means that so long as we can separate logical operators to have disjoint support, we can perform Pauli product measurements of those operators freely. Importantly, the adapters add no further asymptotic overhead beyond that of gauging logical measurements, so if we have $t$ Pauli terms in a logical product measurement we can measure products of them with $\CO(t\omega\log^3\omega)$ space overhead.

\begin{remark}
The gauging measurement can be applied directly to a product of disjoint logical operators, but in the worst case this achieves a less efficient scheme than using adapters. For a product of $t$ terms with maximum representative weight $\omega$, direct gauging logical measurements uses $\CO(t\omega \log^3 t\omega)$ additional qubits, while employing adapters brings the qubit cost to $\CO(t\omega\log^3\omega)$. Although the adapter scheme can also be viewed as a large gauging measurement, it structures the creation of the gauging measurement graph in such a way to reduce the space overhead.
\end{remark}

Unlike brute-force branching, we do not make use of the finer details of gauging logical measurements or universal adapters, and they can be considered black-box procedures for the purpose of our work. See Ref.~\cite{he2025extractors} for a high-level overview of the construction of gauging logical measurements and adapters.

\section{The parallel measurement procedure}\label{sec:parallel_procedure}
In this section we first consider only measuring single logical qubits of the same type ($X$ or $Z$) in parallel on CSS codes. We then incrementally generalise this result to measurements in different bases including $Y$. In Section~\ref{sec:stab_codes} we generalise to stabiliser codes. Then in Section~\ref{sec:pauli_products} we show how to measure Pauli products in parallel, and in Section~\ref{sec:commuting_products} we show how to measure arbitrary commuting sets of Pauli products.

\begin{definition}[Individual Pauli operator]
    Let $Q$ be a quantum LDPC code, and let $\mathcal{B} = (\C^2)^{\otimes k}$ be a choice of tensor decomposition of the logical space $L$ of $Q$ into qubits, i.e. $\mathcal{B} \cong L$. Then an individual Pauli operator acts on a single copy of $\C^2$ in the tensor decomposition.
\end{definition}
In other words, an individual Pauli operator $\bar{P_i}$ acts on only one logical qubit of $Q$. Clearly, this notion is basis-dependent: if a different basis is chosen, the same physical operator on the code could act on multiple (or different) logical qubits.

\begin{definition}[Pauli product]
    Let $Q$ be a quantum LDPC code, and let $\mathcal{B} = (\C^2)^{\otimes k}$ be a choice of tensor decomposition of the logical space $L$ of $Q$ into qubits, i.e. $\mathcal{B} \cong L$. Then a Pauli product may act on multiple copies of $\C^2$ in the tensor decomposition.
\end{definition}
A Pauli product can then be expressed as $\bigotimes_{i \in [k]} \bar{P_i}$ for individual Pauli operators $\bar{P_i}$. Each $\bar{P_i}$ in a Pauli product is called a \textit{Pauli term}, or just \textit{term}, if it is clear from context.

\subsection{Individual logicals of the same type}\label{sec:ind_Z_logs_parallel}

We first consider the simple case where we have many logical qubits in a CSS code to measure simultaneously in the \textit{same} basis, either $Z$ or $X$. For example, measuring $\bar{Z}_1$ and $\bar{Z}_2$ simultaneously. We call these $Z$-\textit{type measurements}, and similarly for $X$-\textit{type measurements}.

\begin{figure}
\[\input{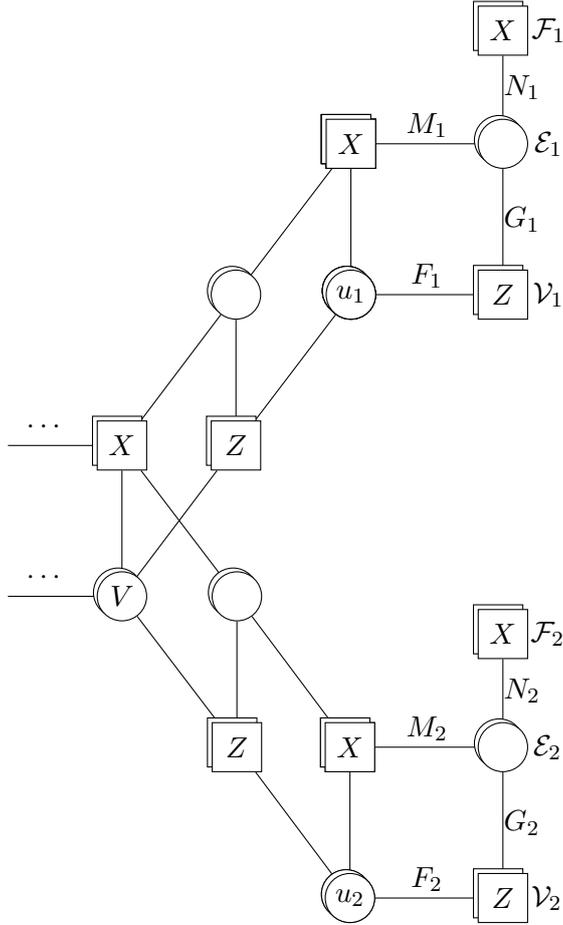}\]
\caption{Branching and gauging measurement for two $\bar{Z}$ logicals. Note that for just two logicals, branching is unnecessary and the gauging measurement systems can be appended in-place, but this scalable Tanner graph generalises to an arbitrary number of overlapping $\bar{Z}$ logicals.}\label{fig:branch_measure1}
\end{figure}

The method first uses brute-force branching, then unlike in Ref.~\cite{ZL} the measurements on leaves are performed using expander graphs~\cite{WY} rather than hypergraph products~\cite{Coh}. 
We illustrate this procedure in Figure~\ref{fig:branch_measure1}.

\begin{lemma}\label{lem:sim_measurement}
Let $Q$ be a CSS quantum LDPC code. We can simultaneously measure individual Pauli operators of the same type ($Z$ or $X$), with representative weight at most $\omega$, on $t$ different logical qubits of $Q$ while retaining fault-distance $d$ and the LDPC property, using $\CO(t\omega (\log t + \log^3\omega))$ new data qubits and checks.
\end{lemma}
\proof
Perform brute-force branching on the logicals $v_i \in \CI$, using the union $V$ of support on data qubits. Next, we use gauging logical measurements, appending expander graphs to the leaves of the brute-force branching tree, measuring each of the individual logicals we desire to measure. Any logicals which we do not wish to measure, but that are contained in $V$, are not fully branched, so are left without a leaf or appended expander graph.

By a combination of Lemmas~\ref{lem:branch_fault_tolerance} and \ref{lem:gauging_fault_tolerance}, this procedure is fault-tolerant with fault-distance $d$, assuming the brute-force branching step and measurement step are each performed for $d$ rounds of stabiliser measurements.

The brute-force branching process requires $\CO(t\omega\log t)$ additional qubits, and the gauging logical measurement a further $\CO(t\omega\log^3\omega)$, resulting in a scaling $\CO\big(t\omega (\log t + \log^3\omega)\big)$.
\endproof

We next consider measuring single logical qubits in parallel, but where the measurement bases are a mixture of $X$s and $Z$s. For this we need to branch both $\bar{X}$ and $\bar{Z}$ operators at the same time.

\subsection{Individual logicals of CSS type}\label{sec:ind_ZX_logs_parallel}

\begin{lemma}\label{lem:ZX_sim_measurement}
    Let $Q$ be a CSS quantum LDPC code. Let the unions of $\bar{Z}$ and $\bar{X}$ logical representatives to be measured have no overlap, i.e. no $\bar{Z}$ representative to be measured intersects with any $\bar{X}$ representative to be measured. We can simultaneously measure individual Pauli operators with representative weight at most $\omega$ of either $Z$ or $X$ type on $t$ different logical qubits on $Q$ while retaining fault-distance and the LDPC property, using $\CO(t\omega(\log t + \log^3\omega))$ new data qubits and checks.
\end{lemma}
\proof
As each $\bar{Z}$ logical has no intersection with an $\bar{X}$ logical, each union of $\bar{Z}$ and $\bar{X}$ logicals can be branched independently, and their representative copies on leaves measured by gauging logical measurement. Formally, we can apply local Cliffords to each $\bar{X}$ logical to convert it into a product of $Z$ Paulis, and compute the branching of all $t$ logicals as though they are $\bar{Z}$ logicals using Lemma~\ref{lem:mixed_branching}, then apply local Cliffords to convert back, to acquire a valid branching for the logicals of mixed types. Hence we can apply Lemma~\ref{lem:sim_measurement}.
\endproof

Unfortunately, representatives of $\bar{Z}$ and $\bar{X}$ operators will generally overlap substantially, even when they commute. When the representatives overlap, the new checks introduced in $Z$-type and $X$-type branches do not commute in general, as the $\bar{X}$ logicals are deformed when branching $\bar{Z}$ logicals and vice versa \cite[App.~F]{ZL}. Thus Lemma~\ref{lem:ZX_sim_measurement} requires the representatives to be chosen such that no $\bar{Z}$ representative to be measured overlaps with any $\bar{X}$ representative to be measured. This is always possible when the logical operators are on different logical qubits, as stated by the following lemma.

\begin{restatable}{lemma}{CSSbasis}\label{lem:rep_basis}
Consider a CSS code \(Q\) with check matrices \(H_X, H_Z\) and parameters \([[n, k, d]]\).
We can find \(\bar{X}\) logical operators \(x_1, \cdots, x_k\in \FF_2^n\) and \(\bar{Z}\) logical operators \( z_1, \cdots, z_k\in \FF_2^n\) of this code, such that \(|x_i\cap z_j| = \delta_{ij}\). Here \(\cap\) denotes intersection as sets. 
\end{restatable}
\begin{corollary}\label{cor:no_overlap}
    In this basis, any set of $\bar{X}$ and $\bar{Z}$ operators acting on different logical qubits have representatives such that their unions have no data qubit overlap.
\end{corollary}

\begin{remark}
    We consider this lemma a standard result for CSS codes and include a proof in Appendix~\ref{app:basis} for completeness. 
    We are aware of independent proofs from~\cite{tjoc2025basis}, see Refs.~\cite[Sec.~4]{shi2025stabilizer} and \cite[Sec.~6]{aasen2025geometrically}. It is also a special case of the `standard form' for stabiliser codes given in Ref.~\cite[Sec.~4.1]{gottesman1997stabilizer}.
\end{remark}

\begin{remark}
    This choice of basis may give logical operators with substantially greater representative weight than the minimal weight possible for each logical operator. Lemma~\ref{lem:rep_basis} gives an upper bound on the representative weight of a $\bar{Z}$ logical of $n-m-k + 1$, so $\omega \in \CO(n)$ for LDPC codes. Similarly, the cleaned $\bar{X}$ logical representatives may have larger weight than otherwise. Thus we emphasise that here $\omega$ is the largest weight of these chosen representatives, not the minimum available for each logical operator.
\end{remark}

Of course, while Lemma~\ref{lem:rep_basis} shows that there exists \textit{a} basis such that the unions of $\bar{Z}$ and $\bar{X}$ representatives do not overlap, there may be \textit{many} such bases, and many valid choices of representatives with different weights; consider for example a surface code patch, which has only 1 logical qubit and many different representatives with different weights satisfying Lemma~\ref{lem:rep_basis}. Thus in practice it will be vital to determine low-weight representatives satisfying the no-overlap condition.

\begin{remark}
For general quantum LDPC codes, finding a logical basis which possesses minimum-weight representatives is NP-hard~\cite{vardy2002intractability}, and hard to approximate~\cite{dumer2003hardness}, even without stipulating any further conditions such as the no-overlap condition above. Therefore, with general surgery schemes for logical measurement, even where the logical basis does not have conditions on its properties, one must be careful and compute the overheads in terms of $\omega$, rather than assuming each representative has weight $d$.
\end{remark}

As an alternative option, rather than choosing new logical operators that satisfy the no-overlap condition, we could instead start with representatives that do overlap. Next, branch the $\bar{Z}$ logicals, and track the overlapping $\bar{X}$ logicals which are deformed by the branching. Then branch these $\bar{X}$ logicals on top of the deformed code, so there will be some connectivity between the $\bar{X}$ and $\bar{Z}$ branches. When the $\bar{X}$ logicals commute with every $\bar{Z}$ logical contained in $V = \bigcup_i v_i$, the deformed $\bar{X}$ logicals will pick up support in the branches with weight proportional to the number of checks required to clean them from $V$. When they do not commute with every contained $\bar{Z}$ logical, which can happen if the logical operator basis is not the chosen basis given by Lemma~\ref{lem:rep_basis}, then they may pick up support on up to $\CO(t\omega \log t)$ new data qubits in the branches. As $\CO(t\omega \log t)$ is not bounded by $\CO(n)$ in general, this can be far worse than choosing a suitable basis, but could be preferable at finite sizes. We discuss this further in Appendix~\ref{app:sim_branching}.

\subsection{Individual logicals of non-CSS type}\label{sec:ind_all_logs_parallel}

We now consider the case when there are also $\bar{Y}$ Paulis to be measured. Unlike in Ref.~\cite{ZL} we do not use ancilla logical qubits or additional codeblocks for this procedure. Instead, we use the same chosen basis as in Section~\ref{sec:ind_ZX_logs_parallel}.

\begin{corollary}
    Let $\bar{\Lambda}_i$ and $\bar{\Lambda}_j$ be any two arbitrary Pauli operators -- $\bar{X}$, $\bar{Y}$ or $\bar{Z}$ -- on different logical qubits $i$, $j$ with representatives given by Lemma~\ref{lem:rep_basis}.
    Let $\alpha$ be an arbitrary data qubit. Either (a) $\bar{\Lambda}_i$ and $\bar{\Lambda}_j$ act on $\alpha$ with the same Pauli operator or (b) at least one of $\bar{\Lambda}_i$ and $\bar{\Lambda}_j$ act with identity on $\alpha$.
\end{corollary}

As a consequence, given any set of $t \leq k$ logical Pauli operator representatives $\bar{\Lambda}_1, \bar{\Lambda}_2,\cdots,\bar{\Lambda}_t$ from Lemma~\ref{lem:rep_basis} on $t$ different logical qubits, these can all be branched such that an equivalent logical operator appears on each leaf.

\begin{lemma}\label{lem:Y_measurements}
    Let $Q$ be a CSS quantum LDPC code, with a chosen basis of logical operators given by Lemma~\ref{lem:rep_basis}. We can simultaneously measure any Pauli operators on $t$ different logical qubits individually on $Q$ while retaining fault-distance and the LDPC property. Letting $\omega$ be the maximum representative weight of logicals to be measured, this uses $\CO\big(t\omega(\log t + \log^3 \omega)\big)$ new data qubits and checks.
\end{lemma}
\proof
First branch using Lemma~\ref{lem:mixed_branching}. After branching, each representative is a product of $Z$ Paulis on each leaf of the tree. Then use gauging logical measurements to measure each Pauli term on each leaf. By a combination of Lemmas \ref{lem:branch_fault_tolerance} and \ref{lem:gauging_fault_tolerance} this operation is fault-tolerant with fault-distance $d$, given that the branching and measurement steps were each performed with at least $d$ rounds of stabiliser measurements.
\endproof

\subsection{Individual logicals on stabiliser codes}\label{sec:stab_codes}

We now show that our protocol for measuring individual logical operators in parallel generalises to stabiliser LDPC codes which are not CSS codes.
For this, we use the `standard form' of stabiliser codes given in Ref.~\cite[Sec.~4.1]{gottesman1997stabilizer}. Recall that in standard form, which can be obtained by Gaussian elimination, the stabiliser generator matrix of an arbitrary stabiliser code $Q$ has the form,
\[
H =\left[H_X | H_Z \right] = \left[
\begin{array}{ccc|ccc}
I & A_1 & A_2 & B & C_1 & C_2 \\
0 & 0 & 0 & D &I & E \\
\end{array}
\right]
\]
where the first three columns are of sizes $r$, $n-k-r$, $k$ respectively, and so are the second three columns; $r$ is the rank of $H_X$ as a matrix over $\F_2$. The two rows are of sizes $r$ and $n-k-r$ respectively.

It is then a standard result that we can pick a basis of logical $\bar{X}$ representatives given by the matrix,
\[
\Xi_{\bar{X}} = \left[
\begin{array}{ccc|ccc}
0 & E^\intercal & I & V_1 & 0 & 0 \\
\end{array}
\right]
\]
where each $\bar{X}$ logical is a row $x_i$ in the matrix, and $V_1 = E^\intercal C_1^\intercal  + C_2^\intercal$. Any two rows $x_i$, $x_j$ act in the same basis on any overlapping data qubits.
This determines a basis of $\bar{Z}$ logical representatives, which up to stabilisers can be described by rows in the matrix,
\[\Xi_{\bar{Z}} = \left[
\begin{array}{ccc|ccc}
0 & 0 & 0 & A_2^\intercal & 0 & I \\
\end{array}
\right].
\]
Similarly, any two rows $z_i$, $z_j$ act in the same basis (the $Z$ basis in this case) on any overlapping data qubits.
Further observe that any two logicals $x_i$ and $z_j$ where $i \neq j$ overlap only on up to $r$ data qubits, in which they act in the same basis. Otherwise, if $i = j$ they anticommute on exactly one data qubit. Hence Cor.~\ref{cor:no_overlap} also applies to this basis, and we acquire the following lemma.

\begin{lemma}\label{lem:stab_measurements}
    Let $Q$ be a stabiliser quantum LDPC code, with a chosen basis of logical operators given by the `standard form'. We can simultaneously measure any individual Pauli operators on $t$ different logical qubits on $Q$ while retaining fault-distance and the LDPC property. Letting $\omega$ be the maximum representative weight of logicals to be measured, this uses $\CO\big(t\omega(\log t + \log^3 \omega)\big)$ new data qubits and checks.
\end{lemma}

We demonstrate the scheme for parallel measurement on stabiliser codes in detail in Appendix~\ref{app:examples_mixed}, using the $\llbracket 8, 3, 3 \rrbracket$ code.

\subsection{Measuring Pauli products}\label{sec:pauli_products}

\begin{figure}\centering
\includegraphics[scale=0.16]
{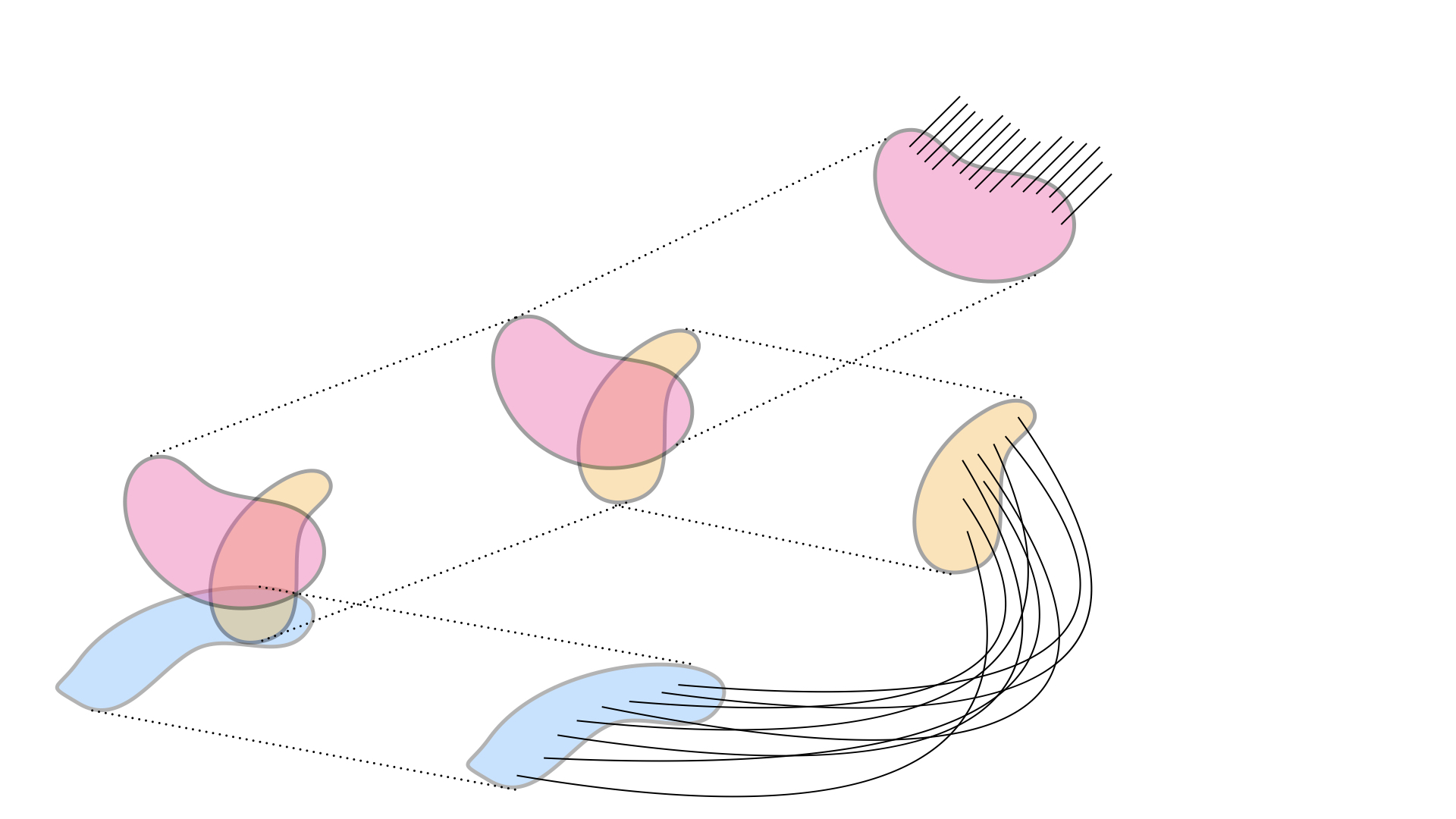}
\caption{Illustration of brute-force branching followed by gauging logical measurements, with adapters for Pauli products.}\label{fig:logical_branching_adapters}
\end{figure}

Measuring logical Pauli products in parallel is a straightforward extension of the scheme outlined above to measure individual logical Paulis in parallel. 
The extension is straightforward because we can separate all logicals included in a set of parallel measurements onto disjoint supports by brute-force branching, and then applying universal adapters to the graphs introduced by the gauging measurement leaves Tanner graph connectivity unaffected elsewhere in the code, see Figure~\ref{fig:logical_branching_adapters}.

\begin{theorem}\label{thm:pauli_products}
Let $Q$ be a stabiliser quantum LDPC code, with a chosen basis of logical operators given by Lemma~\ref{lem:rep_basis}. We can simultaneously measure any logically disjoint Pauli products on $t$ logical qubits on $Q$ while retaining fault-distance and the LDPC property. Letting $\omega$ be the maximum representative weight of logical terms to be measured, this uses $\CO\big(t\omega(\log t + \log^3 \omega)\big)$ new data qubits and checks.
\end{theorem}
\proof
To measure Pauli products, we first repeat the procedure from Lemma~\ref{lem:stab_measurements}, branching each logical operator term in each product and then performing gauging logical measurements.  Now, we also introduce adapters between the measurement graphs for operators whose products are being measured. By Theorem~\ref{thm:adapters} constructing adapters to make a single Pauli product preserves the desiderata of Theorem~\ref{thm:desiderata}, and so fault-distance and the LDPC property. As each logical appears only once in any given product to be measured, by the assumption that we are measuring $t$ terms on different logical qubits, there are at most two adapters connected to any measurement graph, so the LDPC property is preserved for arbitrary sized products. Similarly, each of the Pauli product measurements acts on disjoint sets of logicals, so the fault-distance of Lemma~\ref{lem:gauging_fault_tolerance} carries through. The addition of universal adapters does not change the asymptotic space overhead, so the result follows.
\endproof

Note that $\omega$ is the largest representative weight of a given Pauli term ($\bar{X}$, $\bar{Y}$ or $\bar{Z}$) included in the measurement, not the largest representative weight of a measured Pauli product.

\subsection{Arbitrary commuting products}\label{sec:commuting_products}

The same procedure for measuring logical Pauli products in parallel described above can be extended straightforwardly to the case where we measure logical Pauli products containing terms which have the same, or identity, action on every logical operator.

In this case, the procedure is the same as for Theorem~\ref{thm:pauli_products}, except that when branching individual representatives to leaves we extend the branches (as depicted in Fig.~\ref{fig:summary}d) to give multiple copies of each representative on different leaves. For example, if we wish to measure $\bar{X}_1\bar{Z}_2\bar{I}_3$ simultaneously with $\bar{I}_1\bar{Z}_2\bar{X}_3$, then we have four leaves in total: one for $\bar{X}_1$, two for $\bar{Z}_2$ and one for $\bar{X}_3$. Therefore the space overhead of $\CO\big(T\omega(\log T + \log^3 \omega)\big)$ from Theorem~\ref{thm:pauli_products} carries through, but $T$ is now the number of non-identity terms in the set to be measured.

To measure an arbitrary commuting set $\Theta$ of logical Pauli products in time independent of $t$, we cannot branch, say, $\bar{X}_1$ at the same time as $\bar{Z}_1$ in order to measure $\bar{X}_1\bar{Z}_2\bar{I}_3$ simultaneously with $\bar{Z}_1\bar{X}_2\bar{I}_3$. Therefore we instead use the technique of twist-free surgery from Ref.~\cite{CC} to separate each Pauli product into a pair of commuting products which do not contain $\bar{Y}$ terms, but do have additional support on logical ancilla qubits.

While this is the same approach taken as in Ref.~\cite[App. H]{ZL}, we have two advantages. First, the overall space overhead of our scheme is lower as a consequence of using gauging logical measurements and universal adapters, rather than the CKBB scheme \cite{Coh}. Second, this approach uses ancilla $\ket{Y}$ states, which are potentially costly to prepare in the context of~\cite{ZL}. The authors proposed to either use a qLDPC code that supports a transversal $S$ gate, or distil the $\ket{Y}$ states which incurs an additional overhead. The parallel measurement protocol we have proposed can prepare all required $\ket{Y}$ states in one measurement round. In other words, we bootstrap our measurement of arbitrary commuting sets off the ability of our scheme to measure $Y$ operators without using $\ket{Y}$ states.

\begin{figure}[ht]
    \hfill
     \begin{subfigure}[t]{\textwidth}
     \centering
     \input{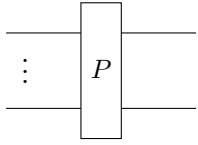}
     \caption{Pauli product measurement $P = iX[u]Z[v]$.}
     \end{subfigure}
     
     \vspace{5mm}
     \begin{subfigure}[t]
     {0.5\textwidth}
         \centering
         \input{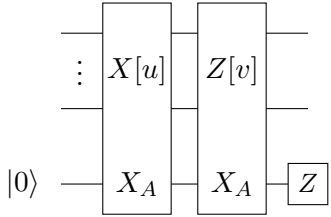}
         \caption{Equivalent measurement for $u\cdot v$ even.}
     \end{subfigure}
     \begin{subfigure}[t]{0.5\textwidth}
     \centering
         \input{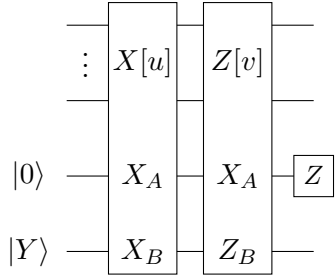}
             \caption{Equivalent measurement for $u\cdot v$ odd.}
     \end{subfigure}
     
     \hfill
     
     \caption{Implementations of the Pauli product measurement $P = i^{u\cdot v}X[u]Z[v]$ from Ref.~\cite{CC}. The measurement outcome of $P$ is determined by the two split Pauli product measurements and the $Z$ measurement. The $\ket{Y}$ state in (c) is left unchanged.}
         \label{fig:twist_free_measures}
\end{figure}

We now describe the procedure to measure arbitrary commuting products in more detail. We use the twist-free gadgets\footnote{The terminology comes from the fact that $Y$ basis measurements can be performed using ``twists'' in surface code patches. We are not using surface codes, but use the same terminology for consistency.} from Ref.~\cite{CC}, which first appeared in Ref.~\cite{KLL}. In such gadgets, we split up a product into two steps using ancillae. This uses the fact that any Pauli product $P$ to be measured can be described as $P = i^{u\cdot v}X[u]Z[v]$ for some binary vectors $u$ and $v$ defining the support of the $X$ and $Z$ components. If there are an even number of $Y$ terms in $P$, so $u \cdot v$ is even, we initialise a single logical $\ket{0}$ state on qubit $A$ and split $P$ into two products: $X[u]X_A$ and $Z[v]X_A$.
If there are an odd number of $Y$ terms, so $u\cdot v$ is odd, we initialize one more logical $\ket{Y} = \frac{1}{\sqrt{2}}(\ket{0} + i\ket{1})$ state on qubit $B$, and split $P$ into $X[u]X_AX_B$ and $Z[v]X_AZ_B$.
This $\ket{Y}$ state serves as a catalyst and is left unchanged by the measurements. These different cases are illustrated in Figure~\ref{fig:twist_free_measures}.

Despite every product in $\Theta$ commuting, the decomposition above may produce products that anticommute. That anticommutation presents a problem for parallel measurements: we would like to measure all the $X$-type products in the same timestep, and then all the $Z$-type products. To ensure this does not happen, the operators in $\Theta$ must satisfy the following condition. 

\begin{definition}\cite[Def.~10]{ZL}
    The operator set $\Theta$ is said to be regular if and only if $u_i\cdot v_j = 0 \mod 2$ for all Pauli products where $i \neq j$ in $\Theta$.
\end{definition}

\begin{example}
    Let $\Theta = \{\bar{X}_1 \otimes \bar{Y}_2, \bar{Z}_1\otimes \bar{X}_2\}$. Then 
    \[\bar{X}_1 \otimes \bar{Y}_2 = \bar{X}\begin{bmatrix}
        1 \\ 1
    \end{bmatrix}\bar{Z}\begin{bmatrix}
        0 \\ 1
    \end{bmatrix},\]
    \[\bar{Z}_1 \otimes \bar{X}_2 = \bar{X}\begin{bmatrix}
        0 \\ 1
    \end{bmatrix}\bar{Z}\begin{bmatrix}
        1 \\ 0
    \end{bmatrix},\]
    ignoring phase factors. In this example $\Theta$ is not regular, as \[\begin{bmatrix}
        1 \\ 1
    \end{bmatrix} \cdot \begin{bmatrix}
        1 \\ 0
    \end{bmatrix} \neq 0, \qquad\begin{bmatrix}
        0 \\ 1
    \end{bmatrix} \cdot \begin{bmatrix}
        1 \\ 1
    \end{bmatrix} \neq 0.\]
    Although $[\bar{X}_1 \otimes \bar{Y}_2, \bar{Z}_1\otimes \bar{X}_2] = 0$, $[\bar{X}_1\otimes \bar{X}_2, \bar{Z}_1\otimes \bar{I}_2] \neq 0$, so the decompositions into $X$ and $Z$-type products do not preserve commutation. As a consequence, the $X$-type products cannot be commuted through the $Z$-type products.
\end{example}

\begin{figure}[ht]
    \hfill
     \begin{subfigure}[t]{\textwidth}
     \centering
     \input{tikz_files/example_commuting1.tikz}
     \caption{Sequential measurement of $Y_1\otimes Y_2\otimes Y_3$ and $Z_1\otimes X_2\otimes Y_3\otimes Y_4$.}
     \end{subfigure}
     
     \vspace{5mm}
     \begin{subfigure}[t]
     {0.5\textwidth}
         \centering
         \input{tikz_files/example_commuting2.tikz}
         \caption{Conversion into twist-free surgeries.}
     \end{subfigure}
     \begin{subfigure}[t]{0.5\textwidth}
     \centering
         \input{tikz_files/example_commuting3.tikz}
             \caption{Commutation of all $X$-type measurements to the beginning.}
     \end{subfigure}
     
     \hfill
     
     \caption{Conversion of the regular set of Pauli product measurements $\Theta = \{Y_1\otimes Y_2\otimes Y_3,Z_1\otimes X_2\otimes Y_3\otimes Y_4\}$ into twist-free surgeries, using the identities in Fig.~\ref{fig:twist_free_measures}. All $X$-type measurements are commuted to the start, which is possible because $\Theta$ is regular. The twist-free surgeries are implemented in two rounds of parallel measurements by surgery, not including the state preparation or final $Z$ measurements. Note that in this case, the set $\Theta$ required only two logical rounds of measurement, as the set has only two Pauli product measurements, and so twist-free surgery does not reduce the time overhead. In general, commuting sets may contain up to $k$ independent Pauli product measurements on $k$ logical qubits.}
         \label{fig:example_commuting}
\end{figure}

Given a general set $\Theta$ of commuting Pauli product measurements, $\Theta$ may not be regular. However, observing that $\Theta$ forms a generating set for a group of Pauli products $\<\Theta\>$, it is shown in Ref.~\cite{ZL} that an alternative generating set $\Theta' \cup \Theta''$ of $\<\Theta\>$ can always be chosen such that the subsets $\Theta'$ and $\Theta''$ are each regular. Measuring $\Theta'$ and then $\Theta''$ then performs the same logical operation on the code as measuring $\Theta$. We omit the details here, but refer the interested reader to \cite[Lem.~9]{ZL}.

As this is a different set of Pauli product operators than $\Theta$, the weights of the Pauli products in the set $\Theta'\cup\Theta''$ may differ substantially from those in $\Theta$.

We then separate the measurement of our Pauli products into 4 steps:
\begin{itemize}
    \item Measure each $X[u_i]X_A$ or $X[u_i]X_AX_B$ for each $X[u_i]$ belonging to $\Theta'$ using a twist-free gadget.
    \item Measure each $Z[v_i]X_A$ or $Z[v_i]X_AZ_B$ for each $Z[v_i]$ belonging to $\Theta'$ using a twist-free gadget.
    \item Measure each $X[u_i]X_A$ or $X[u_i]X_AX_B$ for each $X[u_i]$ belonging to $\Theta''$ using a twist-free gadget.
    \item Measure each $Z[v_i]X_A$ or $Z[v_i]X_AZ_B$ for each $Z[v_i]$ belonging to $\Theta''$ using a twist-free gadget.
\end{itemize}

We perform each step in parallel. Within each step, every Pauli product operator to be measured contains terms which have the same or identity action on every logical qubit. As a consequence, we can use our previous branching and measurement scheme, introducing a leaf for each term to be measured, to perform these measurements in parallel. Each step requires $\CO(t)$ $\ket{0}$ and $\ket{Y}$ ancilla logical states, and $\CO\big(T\omega (\log T + \log^3\omega)\big)$ additional data qubits and checks to measure, where $T$ is the total number of non-identity terms of Pauli products in that step.

This separation into steps using twist-free surgery increases the time overhead of our scheme by a constant factor.
Furthermore, the preparation of $\CO(t)$ $\ket{Y}$ states, which can be reused throughout the computation as they are left unchanged by the twist-free surgery, can be performed in $\CO(d)$ time by measuring $\bar{Y}$ on each of those logical qubits at the start of the computation. 
This removes a major obstacle in Ref.~\cite{ZL}, where the preparation of these $\ket{Y}$ states incurs a significant overhead to distil, or requires transversal $S$ gates.

Finally, we remark that the ancilla logical $\ket{0}$  and $\ket{Y}$ states can be in the same codeblock on which we are performing logical measurements, or in separate codeblocks. The logical measurements involving these states are performed using the same branching and measurement scheme, described at the beginning of this section.
The overhead of preparing these additional high-distance codeblocks is not a major concern, as we only need to prepare them once, which can be done using parallel measurement of individual $\bar{Y}$ logical operators, as specified in Lem.~\ref{lem:Y_measurements}.
Either way, we suffer a rate loss of at most a factor of $3$, which is acceptable in the asymptotic regime.

\section{Discussion}

In this work we have presented a scheme for parallel quantum LDPC code surgery to fault-tolerantly measure logically disjoint Pauli products supported on $t$ logical qubits using $\CO\big(t\omega(\log t+\log^3 \omega)\big)$ ancilla data qubits, and in time $\CO(d)$, independent of $t$. This scheme requires no ancilla logical qubits. We then showed how to leverage this scheme to bootstrap a measurement protocol for arbitrary commuting sets of Pauli products, using $\CO(t)$ ancilla logical qubits, and still not exceeding $\CO(d)$ time to measure the commuting set.

Throughout, we assumed that branching and gauging measurement are separate steps, performed independently, to aid with the proofs of fault-distance in Appendix~\ref{app:ft_proofs}. We expect that they can be performed simultaneously while maintaining the fault-distance, but have not proved this.

In practice, for small codes it has been found that decongestion is often not necessary, so the $\log^3\omega$ factor vanishes from the overhead~\cite{IGND, WY}, and in general the upper bounds on overhead to preserve the distance tend to be very loose in quantum LDPC surgery~\cite{Cow24}. Moreover, recent work implies that the $\log^3\omega$ factor can be reduced to $\log \omega$~\cite{hsieh2025simplified}, although the precise details are yet to be worked out. The discussion in Ref.~\cite[Sec.~7.1.2]{hsieh2025simplified} also gives evidence that our scheme may have optimal space overhead for a generic parallel logical measurement scheme, up to polylog factors, as our scheme can be seen as weight reduction: first of the checks required to measure the logical operators (gauging logical measurement), then of the qubits that those checks overlap on (brute-force branching).

It would be useful to avoid brute-force branching, to remove the $\log t$ factor and the formidable constants associated with branches, possibly by combining the gauging logical measurements framework with devised sticking~\cite{ZL}, rather than brute-force branching. Unfortunately we could not find a construction for doing so which allowed us to provably use expander (hyper)graphs with gauged cycles, and so retain a space overhead sub-quadratic in $d$. Ref.~\cite{zheng2025highrate} takes a similar approach, albeit without provable bounds on overhead. We remark that the use of the logical basis in standard form, along with toggling of connections using single-qubit Cliffords as in Lem.~\ref{lem:mixed_branching}, allow for a different measurement scheme for logically disjoint Pauli products by devised sticking in $\CO(\min(n d, t\omega d))$ space, an improvement on the $\CO(t^2\omega d)$ of Ref.~\cite{ZL}. This would require moving to a subsystem code, however. We leave the details for future work. 

If parallel LDPC code surgery is to be performed with branches, it is imperative to minimise the weights of logical representatives while maintaining suitable overlap properties. 
Another desirable improvement is to reduce the $\CO(d)$ time overhead~\cite{HDTV} of a logical round while retaining the parallelism of our scheme, without increasing space overhead substantially. Recent work on fast surgery accomplishes the reduction to $\CO(1)$ syndrome measurement rounds, but either at the expense of space~\cite{baspin2025fast} or addressability on logical qubits~\cite{cowtan2025fast}. Interestingly, the addressable scheme of Ref.~\cite{baspin2025fast} has the same asymptotic spacetime overhead as the parallel measurement scheme in this work, suggesting that there may be fundamental bounds on the asymptotic spacetime volume of distance-preserving addressable logical measurements on general quantum LDPC codes.

Recent work has focussed on performing surgery on quantum LDPC codes in a fixed-connectivity setting~\cite{he2025extractors,yoder2025tour}. There are aspects of the scheme presented in this work that function in a fixed-connectivity setting. For example, branches for each single-qubit individual logical Pauli operator could be constructed in advance, along with their expander graphs for gauging logical measurement, then branches deleted when required. However, the space overhead of doing this will be exorbitant, and including all the connectivity for all possible adapters even more so. How to attain a reasonable degree of parallelism in a fixed-connectivity regime, either at blocklengths of practical relevance for large quantum algorithms or in the asymptotic regime, without increasing space cost substantially remains an open question, which we leave for future work.

The most important outstanding question for practical purposes is how much effect the combination of brute-force branching and measurement has on the circuit-level threshold of the code. Discerning this requires substantial circuit-level noise simulations and will depend heavily on the syndrome extraction circuits and decoder, so we also defer this to future work.

\section{Acknowledgements}
We used TikZit~\cite{TkZ} to generate the Tanner graph figures in this paper.
This work was partially completed while A.C.~was employed at Quantinuum. A.C.~is grateful for their support and helpful discussions with colleagues there. A.C. is also grateful to Ben Ide for pointing out the standard form of logical representatives for stabiliser codes in Ref.~\cite{gottesman1997stabilizer}.
Z.H.~is supported by the MIT Department of Mathematics and the NSF Graduate Research Fellowship Program under Grant No. 2141064. 
D.W.~is currently on leave from The University of Sydney and is employed by PsiQuantum. 
Part of this work was initiated when Z.H., A.C.~and D.W.~were visiting the Fault-Tolerant Quantum Technologies Workshop in Benasque, 2024. 
We are grateful to the organisers of the workshop.

\bibliographystyle{unsrt}
\bibliography{bibtex}

\input{appendix}
\end{document}

%% file: basic.tikzstyles
\tikzstyle{boundary vertex}=[inner sep=0mm, minimum size=1mm, shape=circle, draw=black, fill=black]
\tikzstyle{grey_dot}=[fill={rgb,255: red,191; green,191; blue,191}, draw={rgb,255: red,191; green,191; blue,191}, shape=circle, minimum size=1mm, inner sep=0mm]
\tikzstyle{blue_dot}=[fill={rgb,255: red,202; green,251; blue,255}, draw=black, shape=circle, minimum size=1.5mm, inner sep=0mm]
\tikzstyle{white_dot}=[fill=white, draw=black, shape=circle, minimum size=1.5mm, inner sep=0mm]
\tikzstyle{red_dot}=[fill=red, draw=black, shape=circle, minimum size=1.5mm, inner sep=0mm]
\tikzstyle{box}=[fill=white, draw=black, shape=rectangle]
\tikzstyle{scalable_dot}=[fill=white, draw=black, shape=circle, minimum size=6.5mm, inner sep=0mm]
\tikzstyle{scalable_box}=[fill=white, draw=black, shape=rectangle, minimum height=6.5mm, minimum width=6.5mm]
\tikzstyle{scalable_meta_Z}=[fill={rgb,255: red,255; green,170; blue,162}, draw=black, shape=diamond, minimum height=8mm, minimum width=8mm]

\tikzstyle{arrow}=[->]
\tikzstyle{red_arrow}=[->, draw=red]
\tikzstyle{cyan_arrow}=[->, draw=cyan]
\tikzstyle{red_dash}=[-, dashed, draw=red]
\tikzstyle{grey dash}=[-, fill=none, draw={rgb,255: red,191; green,191; blue,191}, dashed]
\tikzstyle{dashed arrow}=[->, dashed]
\tikzstyle{red_edge}=[-, draw=red]
\tikzstyle{green_edge}=[-, fill=none, draw=green]
\tikzstyle{green_dash}=[-, draw=green, dashed]
\tikzstyle{orange_edge}=[-, draw={rgb,255: red,255; green,183; blue,0}, fill=none, ultra thick]
\tikzstyle{pink_edge}=[-, draw={rgb,255: red,255; green,83; blue,227}, ultra thick]
\tikzstyle{blue_edge}=[-, draw={rgb,255: red,43; green,181; blue,255}, ultra thick]

%% file: quantum.tikzstyles
\tikzstyle{gate}=[shape=rectangle, text height=1.5ex, text depth=0.25ex, yshift=0.5mm, fill=white, draw=black, minimum height=5mm, yshift=-0.5mm, minimum width=5mm, font={\small}, tikzit category=circuit]
\tikzstyle{big gate}=[shape=rectangle, text height=1.5ex, text depth=0.25ex, yshift=0.5mm, fill=white, draw=black, minimum height=18mm, yshift=-0.5mm, minimum width=5mm, font={\small}, tikzit category=circuit]
\tikzstyle{Z dot}=[inner sep=0mm, minimum size=2mm, shape=circle, draw=black, fill={rgb,255: red,221; green,255; blue,221}, tikzit category=zx]
\tikzstyle{Z phase dot}=[minimum size=5mm, font={\footnotesize\boldmath}, shape=rectangle, rounded corners=2mm, inner sep=0.2mm, outer sep=-2mm, scale=0.8, tikzit shape=circle, draw=black, fill={rgb,255: red,221; green,255; blue,221}, tikzit draw=blue, tikzit category=zx]
\tikzstyle{X dot}=[Z dot, shape=circle, draw=black, fill={rgb,255: red,255; green,136; blue,136}, tikzit category=zx]
\tikzstyle{X phase dot}=[Z phase dot, tikzit shape=circle, tikzit draw=blue, fill={rgb,255: red,255; green,136; blue,136}, font={\footnotesize\boldmath}, tikzit category=zx]
\tikzstyle{hadamard}=[fill=yellow, draw=black, shape=rectangle, inner sep=0.6mm, minimum height=1.5mm, minimum width=1.5mm, tikzit category=zx]
\tikzstyle{paulibox}=[fill={rgb,255: red,221; green,221; blue,255}, draw=black, shape=rectangle, inner sep=0.6mm, minimum height=5mm, minimum width=5mm, font={\footnotesize}, text height=1.5ex, text depth=0.25ex, tikzit category=zx]
\tikzstyle{vertex}=[inner sep=0mm, minimum size=1mm, shape=circle, draw=black, fill=black, tikzit category=misc]
\tikzstyle{vertex set}=[inner sep=0mm, minimum size=1mm, shape=circle, draw=black, fill=white, font={\footnotesize\boldmath}, tikzit category=misc]
\tikzstyle{small black dot}=[fill=black, draw=black, shape=circle, inner sep=0pt, minimum width=1.2mm, tikzit category=circuit]
\tikzstyle{cnot ctrl}=[fill=black, draw=black, shape=circle, inner sep=0pt, minimum width=1.2mm, tikzit category=circuit]
\tikzstyle{cnot targ}=[fill=white, draw=white, shape=circle, tikzit category=circuit, label={center:$\oplus$}, inner sep=0pt, minimum width=2.1mm, tikzit fill={rgb,255: red,102; green,204; blue,255}, tikzit draw=black]
\tikzstyle{ket}=[fill=white, draw=black, shape=regular polygon, regular polygon sides=3, regular polygon rotate=-30, scale=0.7, inner sep=1pt, tikzit category=circuit, tikzit shape=rectangle, tikzit fill=green]
\tikzstyle{bra}=[fill=white, draw=black, shape=regular polygon, regular polygon sides=3, regular polygon rotate=30, scale=0.7, inner sep=1pt, tikzit category=circuit, tikzit shape=rectangle, tikzit fill=red]
\tikzstyle{scalar}=[shape=rectangle, text height=1.5ex, text depth=0.25ex, yshift=0.5mm, fill=white, draw=black, minimum height=5mm, yshift=-0.5mm, minimum width=5mm, font={\small}]
\tikzstyle{clabel}=[fill=white, draw=none, shape=rectangle, tikzit fill={rgb,255: red,56; green,255; blue,242}, font={\footnotesize}, inner sep=1pt, tikzit category=labels]
\tikzstyle{empty diagram}=[draw={gray!40!white}, dashed, shape=rectangle, minimum width=1cm, minimum height=1cm, tikzit category=misc]
\tikzstyle{bigger gate}=[fill=white, draw=black, shape=rectangle, minimum height=28mm, minimum width=5mm, font={\small}, tikzit category=circuit]
\tikzstyle{wider gate}=[fill=white, draw=black, shape=rectangle, minimum height=28mm, minimum width=9mm, font={\small}]
\tikzstyle{biggest gate}=[fill=white, draw=black, shape=rectangle, minimum height=36mm, minimum width=9mm, font={\small}]
\tikzstyle{even biggest gate}=[fill=white, draw=black, shape=rectangle, minimum height=46mm, minimum width=9mm, font={\small}]

\tikzstyle{hadamard edge}=[-, dashed, dash pattern=on 2pt off 0.5pt, thick, draw={rgb,255: red,68; green,136; blue,255}]
\tikzstyle{box edge}=[-, dashed, dash pattern=on 2pt off 0.5pt, thick, draw={rgb,255: red,203; green,192; blue,225}]
\tikzstyle{brace edge}=[-, tikzit draw=blue, decorate, decoration={brace,amplitude=1mm,raise=-1mm}]
\tikzstyle{diredge}=[->]
\tikzstyle{double edge}=[-, double, shorten <=-1mm, shorten >=-1mm, double distance=2pt]
\tikzstyle{gray edge}=[-, {gray!60!white}]
\tikzstyle{pointer edge}=[->, very thick, gray]
\tikzstyle{boldedge}=[-, line width=1.6pt, shorten <=-0.17mm, shorten >=-0.17mm]

%% file: appendix.tex
\appendix

\section{Connections to homological algebra}
\subsection{Hypergraph products}
\label{app:hypergraph_prods}

Hypergraph products~\cite{TZ} are essentially tensor products of chain complexes, where one of the complexes has been dualised. That is, denoting the hypergraph product $[C, D]_\bullet$\footnote{This notation comes from the fact that the hypergraph product is an internal hom $[-,-]$ in the compact closed category of finite-dimensional bounded chain complexes over $\F_2$ \cite{Mac}.} then
\[[C, D]_\bullet := (D\otimes C^*)_\bullet\]
where $-\otimes-$ is the tensor product of chain complexes \cite{AC, Weib}.

Given $C_\bullet$ and $D_\bullet$ as scalable Tanner graphs,
\[\input{tikz_files/tensor_prod1.tikz}\]
which are classical codes and so have no $Z$ checks (in our convention), then the hypergraph product $[C, D]_\bullet$ is
\[\input{tikz_files/hyper_prod_CD.tikz}\]
in scalable Tanner graph notation.

In particular, if we take the repetition code $R_\bullet = R_1 \rightarrow R_0$, with parity-check matrix
\[\begin{pmatrix}
1 & 1 & 0 & \cdots & 0\\
0 & 1 & 1 & \cdots & 0\\
0 & 0 & 1 & \cdots & 0\\
\vdots & \vdots & \vdots & \ddots & \vdots\\
0 & 0 & 0 & \cdots & 1
\end{pmatrix}\]
then the scalable Tanner graph of $[R,D]_\bullet$ is
\[\input{tikz_files/hyper_prod_CD2.tikz}\]
where all horizontal edges carry the label $I$, i.e.~they are identity maps. We can see that if $D_\bullet$ is the classical code corresponding to a $\bar{Z}$ logical in a larger code $Q$, and its incident $X$ checks, then gluing $[R,D]_\bullet$ into $Q$ as a branching sticker makes another equivalent logical on each odd layer of this scalable Tanner graph; in particular there is an equivalent logical on the very last layer. In brute-force branching, we initialise trees of such hypergraph products where each branch uses $R_\bullet$ with the check matrix $\begin{pmatrix} 1 &1 \end{pmatrix}$, but varying $D_\bullet$ to have a classical code decreasing in size, to give equivalent logicals without overlap, as in Fig.~\ref{fig:all_branching}.

\subsection{Branching as a series of mapping cylinders}\label{app:branch_cylinders}

When the initial code $Q$ is CSS, brute-force branching of all $Z$ or $X$-type operators can be rephrased in the language of chain complexes. This subsection assumes the reader is familiar with elementary homological algebra. By standard theory~\cite{BE2}, $Q$ can be viewed as a chain complex $C_\bullet$ over $\F_2$:

\[C_\bullet = \begin{tikzcd}
    C_2 \arrow[r, "\del_2"] & C_1 \arrow[r, "\del_1"] & C_0,
\end{tikzcd}\]
where $\del_2 = H_Z^T$ and $\del_1 = H_X$, and each vector space is assigned a basis, such that 
\[C_2 = \F_2^{m_z},\qquad C_1 = \F_2^n, \qquad C_0 = \F_2^{m_x},\]
where $n$, $m_x$ and $m_z$ are the number of data qubits, $X$-type and $Z$-type stabiliser generators respectively. Each data qubit in $Q$ can be identified with a basis element in $C_1$, and $X$($Z$)-checks can be identified with basis elements in $C_0$($C_2$) respectively.
The number of logical qubits in $Q$ is determined by the homology of $C_\bullet$ at degree 1, i.e.
\[k_Q = \dim H_1(C).\]
Furthermore, each $\bar{Z}$ logical operator can be identified with a vector $v\in C_1$, such that $v\in \ker \del_1$; for each nontrivial $\bar{Z}$ logical operator, $v \in \ker \del_1 \backslash \im \del_2$. Each equivalence class of operators (up to $Z$ stabilisers) with representative $\bar{Z}$ is associated to $[v]$, the equivalence class in $H_1(C) = \ker \del_1/\im \del_2$.

Let $\{\bar{Z}_i\}_{i \in I}$ be a set of $\bar{Z}$ logical operators to branch. $\{\bar{Z}_i\}_{i \in I}$ can be viewed as a set of $\F_2$-vectors $v_i \in C_1$.

Let $A_1 \subset C_1$ be a subspace of $C_1$, defined by
$A_1 = \bigcup_i\mathrm{supp}(v_i)$. That is, a basis element of $C_1$ is included in $A_1$ if and only if it is in the support of a $\bar{Z}$ operator to be branched.

Then define a matrix $\del_1^A := \del_1 \restriction_{A_1}$, restricting $H_X$ to its support on $A_1$. The codomain of this matrix is $A_0 = \bigcup_{u \in \im \del_1^A} \mathrm{supp}(u)$. In other words, $A_0$ has as basis elements all $X$ checks which are incident to any qubits in the support of any $\bar{Z}$ operators to be branched.

The complex $A_\bullet = A_1 \rightarrow A_0$ is a subcomplex of $C_\bullet$, i.e. there is an injective chain map $f_\bullet: A_\bullet \hookrightarrow C_\bullet$, which maps basis elements to basis elements. As a commutative diagram,
\[\begin{tikzcd}
    0 \arrow[r] \arrow[d, "f_2", hookrightarrow] & A_1 \arrow[d, "f_1", hookrightarrow]\arrow[r, "\del_1^A"] & A_0\arrow[d, "f_0", hookrightarrow] \\
    C_2 \arrow[r, "\del_2"] & C_1 \arrow[r, "\del_1"] & C_0
\end{tikzcd}.\]
Note that the zero object and $f_2$ map can be omitted, as the map must always be zero. As $f_1$ and $f_0$ map basis elements to basis elements, each of them has the form
\[\begin{pmatrix}
    I \\ 0
\end{pmatrix}\]
up to permutation.

The branch can then be constructed as $\mathrm{cyl}(f)$, where $\mathrm{cyl}(\cdot)$ is the mapping cylinder~\cite[Sec.~1.5]{Weib}. Diagrammatically,
\[\mathrm{cyl}(f) = \begin{tikzcd}
    & A_1 \arrow[r, "\del_1^A"] & A_0 \\
    A_1 \arrow[ur, "I"]\arrow[dr, "f_1"] \arrow[r, "\del_1^A"] & A_0 \arrow[ur, "I"] \arrow[dr, "f_0"]& \\
    C_2 \arrow[r, "\del_2"] & C_1 \arrow[r, "\del_1"] & C_0
\end{tikzcd}\]
where direct sums are taken vertically on the vector spaces.

To orient this construction in terms of quantum codes and Sec.~\ref{sec:branching}, observe that for any $\bar{Z}_i$ with $i \in I$, multiplying by the appropriate element in $\im \del_1^{\mathrm{cyl}(A)}$ will shift the support $v_i$ from its location in the original code to the new data qubits $A_1$ on the top row. That is, there is now an equivalent representative of $\bar{Z}_i$ with support wholly outside of the original code, and so the mapping cylinder creates a branch.

By inspection there is an injective chain map $\alpha_\bullet : C_\bullet \hookrightarrow \mathrm{cyl}(f)$, mapping $C_i \rightarrow C_i$.
\begin{lemma}[\cite{Weib}, Lem.~1.5.6]
    The inclusion $\alpha_\bullet : C_\bullet \hookrightarrow \mathrm{cyl}(f)$ is a quasi-isomorphism.
    \label{lem:mapping_cyl}
\end{lemma}
That is, the restriction and corestriction of $\alpha_\bullet$ to homology spaces is an isomorphism, i.e. $H_i(\alpha)$ is an isomorphism for all degrees $i$.

As a consequence, the addition of a branch introduces no new logical operators. Note that the above Lemma is stronger than $H_1(C) \cong H_1(\mathrm{cyl}(f))$ as vector spaces; it also dictates that this isomorphism is given by $H_1(\alpha)$, which means that every $\bar{Z}$ logical operator in $C_\bullet$ is also present in $\mathrm{cyl}(f)$ and the equivalence classes are preserved, so it is not the case that some logical qubits are removed and replaced with different ones. See e.g. Ref.~\cite{CowBu} or Ref.~\cite[Sec.~2.2.3]{cowtan2025homology} for discussion on interpreting logical maps, i.e. maps between homology spaces, from physical maps, i.e. maps between chain complexes.

Next, one can make a branch for a different set $\{\bar{Z}_j\}_{j \in J}$ of $\bar{Z}$ logical operators, in the same manner.\footnote{Indeed, $I \cap J$ does not need to be zero, which allows for branching the same logical operator to multiple different leaves.} This second branch is still well-defined by checking the diagram,

\[\mathrm{cyl}(f+g) = \begin{tikzcd}
    & A_1 \arrow[r, "\del_1^A"] & A_0 \\
    A_1 \arrow[ur, "I"]\arrow[dr, "f_1"] \arrow[r, "\del_1^A"] & A_0 \arrow[ur, "I"] \arrow[dr, "f_0"]& \\
    C_2 \arrow[r, "\del_2"] & C_1 \arrow[r, "\del_1"] & C_0 \\
    B_1 \arrow[dr, "I"]\arrow[ur, "g_1"] \arrow[r, "\del_1^B"] & B_0 \arrow[dr, "I"] \arrow[ur, "g_0"]& \\
    & B_1 \arrow[r, "\del_1^B"] & B_0
\end{tikzcd}\]

By iteration, one can build up a series of mapping cylinders on $A_\bullet$ and $B_\bullet$, branching all $\bar{Z}$ logical operators to have leaves. By applying Lemma~\ref{lem:mapping_cyl} sequentially, one immediately acquires a series of quasi-isomorphisms,
\[C_\bullet\  \tilde{\rightarrow} \ \mathrm{cyl}(f) \ \tilde{\rightarrow} \ \mathrm{cyl}(f+g) \ \tilde{\rightarrow} \ \cdots\]
and so brute-force branching introduces no new logical operators. This yields an alternative, more algebraic, proof of Lemma~\ref{lem:branch_no_new}.

\begin{remark}
    In Ref.~\cite[App.~A]{ide2024fault} mapping cylinders were suggested as a potential formalism for treating joint measurement. Here we use them for branching, a different but related application.
\end{remark}

\begin{remark}
    We claim that the same proof via mapping cylinders also applies to non-CSS stabiliser codes, but this requires including the symplectic form in the chain complexes explicitly~\cite{pesah2025fault}, which complicates the construction. In our case, all the branching on non-CSS codes toggles check type to make the code locally CSS, see Lemma~\ref{lem:mixed_branching}, which obviates the need for such a formalism for branching, as Lemma~\ref{lem:branch_no_new} relies only on the code locally around the representatives being branched.
\end{remark}

\section{Toy examples of brute-force branching}\label{app:example_branching}
In this appendix we give explicit examples of brute-force branching and measurement.

\subsection{Brute-force branching $\bar{Z}$ logicals}
We start with a toy set of $\bar{Z}$ logicals. Consider a CSS code $Q$, with Tanner graph $\mathcal{T}(Q)$, such that there is a Tanner subgraph $\mathcal{G}(V_{\rm data}\cup V_{\rm check}, E)$,
\[\input{tikz_files/branching_example1.tikz}\]
For simplicity of presentation, we do not consider the rest of the Tanner graph $\mathcal{T}(Q)$ (and therefore do not specify the actual initial code $Q$). Crucially, the $X$-checks in $\mathcal{G}$ are the \textit{only} $X$-checks incident to the data qubits in $V_{\rm data}$; otherwise, one can check that the deformed code would not generally have commuting stabilisers. The data qubits in $V_{\rm data}$ may be incident to $Z$-checks elsewhere in $\mathcal{T}(Q)$, and the $X$-checks in $V_{\rm check}$ may be incident to other data qubits in $\mathcal{T}(Q)$.

Observe that there are 3 independent $\bar{Z}$ logicals in $\mathcal{G}$: $v_1$, $v_2$ and $v_3$. We highlight $v_1$ in orange,
\[\input{tikz_files/branching_example1_orange.tikz}\]
$v_2$ in pink,
\[\input{tikz_files/branching_example1_pink.tikz}\]
and $v_3$ in blue.
\[\input{tikz_files/branching_example1_blue.tikz}\]
Each of the $\bar{Z}$ logicals in $\mathcal{G}$ have weight at least 3, including those that are linear combinations.
Following the description in Sec.~\ref{sec:branching}, let $\CI = \{v_1, v_2, v_3\}$, and $t=3$. We choose $\lceil \frac{t}{2}\rceil = 2$ logicals from $\CI$, say $v_1$ and $v_2$. We say that $G_1(S_1\cup 
\chi_1, E_1)$ is the subgraph of $\mathcal{G}$ where $S_1 = {\rm supp}(v_1) \cup {\rm supp}(v_2)$ and $\chi_1$ is the set of $X$-checks incident to any vertex in $S_1$. Then the branch for the subgraph $G_1$ has the following form:
\[\input{tikz_files/branching_example2.tikz}\]
Call this Tanner graph $B_1(\mathcal{G})$. Each data qubit in the support of $S_1$ has a new $Z$-check in $S_1^\intercal$ connected to it; each $X$-check in $\chi_1$ has a new data qubit in $\chi_1^\intercal$ connected to it, and these have connections between them which mirror $G_1$. Then, each new $Z$-check in $S_1^\intercal$ has another new data qubit connected to it, and each new data qubit in $\chi_1^\intercal$ has another new $X$-check connected to it, in $H_1$, the copy of $G_1$, which is placed on the top left. Note that there is an injection of graphs $\mathcal{G} \subset B_1(\mathcal{G})$, i.e. we have built on top of the previous subgraph $\mathcal{G}$.
In $B_1(\mathcal{G})$ each of the logicals $v_1, v_2$ can be cleaned to the end of the branch.

$v_1'$, which is $v_1$ after cleaning, is highlighted in orange below,
\[\input{tikz_files/branching_example2_orange.tikz}\]
and $v_2'$ is highlighted in pink:
\[\input{tikz_files/branching_example2_pink.tikz}\]
Now, $v_1'$ and $v_2'$ are not disjoint from one another, but are disjoint from $v_3$:
\[\input{tikz_files/branching_example2_blue.tikz}\]
As $v_3$, the other $\lfloor \frac{t}{2}\rfloor = 1$ operator, is now disjoint from $v_1'$ and $v_2'$, we could leave $v_3$ unchanged, but it is illustrative to branch $v_3$ as well, to indicate the structure for $t >3 $. Repeating the branching procedure for $v_3$ yields the following Tanner graph:
\[\input{tikz_files/branching_example3.tikz}\]
where $v_3'$ is highlighted in blue below:
\[\input{tikz_files/branching_example3_blue.tikz}\]
One could then continue, branching out $v_1'$ and $v_2'$, which we omit.
This explicit Tanner graph is an instantiation of the more abstract illustration in Fig.~\ref{fig:logical_branching} before the last round of branching, which would yield disjoint representatives $v_1'', v_2''$.

Immediately from this toy example, one can see that for low-distance or low-rate codes, branching may incur a substantial practical cost. However, for codes which are high-distance and high-rate, branching followed by gauging measurement and adapters is asymptotically efficient.

\subsection{Parallel measurement of mixed logicals}\label{app:examples_mixed}
We now give a toy example of brute-force branching and measuring mixed logicals. For this example, we use the $\llbracket 8, 3, 3\rrbracket$ non-CSS stabiliser code from Ref.~\cite[Table~3.3]{gottesman1997stabilizer}.

The $\llbracket 8, 3, 3 \rrbracket$ code has the stabiliser matrix,
\[
\left[
\begin{array}{cccccccc|cccccccc}
1 & 1 & 1 & 1 & 1 & 1 & 1 & 1 &  0 & 0 & 0 & 0 & 0 & 0 & 0 & 0 \\
0 & 0 & 0 & 0 & 0 & 0 & 0 & 0 &  1 & 1 & 1 & 1 & 1 & 1 & 1 & 1\\
0 & 1 & 0 & 1 & 1 & 0 & 1 & 0 &  0 & 0 & 0 & 0 & 1 & 1 & 1 & 1 \\
0 & 1 & 0 & 1 & 0 & 1 & 0 & 1 &  0 & 0 & 1 & 1 & 0 & 0 & 1 & 1 \\
0 & 1 & 1 & 0 & 1 & 0 & 0 & 1 &  0 & 1 & 0 & 1 & 0 & 1 & 0 & 1 \\
\end{array}
\right].
\]
As there are checks with weight $8$, the same as the length of the code, there is not a great deal of point in performing LDPC surgery on the $\llbracket 8, 3, 3 \rrbracket$ code, as there is no sparsity to be preserved. Nevertheless, it is a reasonable toy example for the purpose of demonstration.

In standard form~\cite[Sec.~4.1]{gottesman1997stabilizer}, the logical operators have the structure:

\[
\left[
\begin{array}{c|cccccccc|cccccccc}
\bar{X}_1 & 1 & 1 & 0 & 0 & 0 & 0 & 0 & 0 &  0 & 0 & 0 & 0 & 0 & 1 & 0 & 1 \\
\bar{X}_2 & 1 & 0 & 1 & 0 & 0 & 0 & 0 & 0 &  0 & 0 & 0 & 1 & 0 & 0 & 1 & 0 \\
\bar{X}_3 & 1 & 0 & 0 & 0 & 1 & 0 & 0 & 0 &  0 & 0 & 0 & 1 & 0 & 1 & 0 & 0 \\
\bar{Z}_1 & 0 & 0 & 0 & 0 & 0 & 0 & 0 & 0 &  0 & 1 & 0 & 1 & 0 & 1 & 0 & 1 \\
\bar{Z}_2 & 0 & 0 & 0 & 0 & 0 & 0 & 0 & 0 &  0 & 0 & 1 & 1 & 0 & 0 & 1 & 1 \\
\bar{Z}_3 & 0 & 0 & 0 & 0 & 0 & 0 & 0 & 0 &  0 & 0 & 0 & 0 & 1 & 1 & 1 & 1 \\
\end{array}
\right].
\]

We choose an arbitrary selection of logical operators on disjoint logical qubits to measure in parallel: $\bar{X}_1, \bar{Z}_2, \bar{Y}_3$. As they act on disjoint logical qubits, and the logical basis is in standard form, by Lemma~\ref{lem:mixed_branching} we can branch simultaneously.

For illustrative purposes, we shall partition the terms into $\{\bar{X}_1$, $\bar{Z}_2\}$ and $\{\bar{Y}_3\}$, and branch $\{\bar{X}_1$, $\bar{Z}_2\}$ together, then $\{\bar{Y}_3\}$. Then we shall split $\{\bar{X}_1$, $\bar{Z}_2\}$ onto leaves, and finally apply the gauging logical measurement scheme to measure $\bar{X}_1, \bar{Z}_2, \bar{Y}_3$.

The union of $\bar{X}_1$ and $\bar{Z}_2$ acts as
\[
\left[
\begin{array}{c|cccccccc|cccccccc}
\bar{X}_1 \cup \bar{Z}_2 & 1 & 1 & 0 & 0 & 0 & 0 & 0 & 0 &  0 & 0 & 1 & 1 & 0 & 1 & 1 & 1 \\
\end{array}
\right].
\]
Every check in the code has a component which anticommutes with the action of $\bar{X}_1 \cup \bar{Z}_2$, so the first branch will introduce 5 new data qubits, as well as the 7 new checks. The local structure of the Tanner graph $\mathcal{T}(Q)$ around $\bar{X}_1 \cup \bar{Z}_2$ is
\[\input{tikz_files/local_tannergraph_X1Z2_1.tikz}\]
where vertices are labelled according to columns and rows in the check matrix, and edges are labelled according to Def.~\ref{def:tanner_graph}.

$\bar{X}_1$ is highlighted in orange,
\[\input{tikz_files/local_tannergraph_X1Z2_1_orange.tikz}\]
and $\bar{Z}_2$ in pink:
\[\input{tikz_files/local_tannergraph_X1Z2_1_pink.tikz}\]

As the $X$-component of $\bar{X}_1$ is disjoint from the $Z$-component of $\bar{Z}_2$, and vice versa, we can toggle the $X$-component of the union, on qubits 1 \& 2, to act as $Z$ instead, using single-qubit Cliffords (in this case Hadamards), which yields the local Tanner graph,
\[\input{tikz_files/local_tannergraph_X1Z2_1_permuted.tikz}\]
In other words, despite the whole code not being CSS, we can consider the local code around the logicals to be branched to be CSS, and define a subcomplex as in App.~\ref{app:branch_cylinders}.

Using Tanner graphs, the branch has the following form:
\[\input{tikz_files/local_tannergraph_X1Z2_1_branch.tikz}\]
Afterwards, to make the local code agree with the rest of $Q$ again, we toggle the $X$-component of the union in $Q$ back.
\[\input{tikz_files/local_tannergraph_X1Z2_1_permuteback.tikz}\]
Observe that after cleaning to the end of the branch, both $\bar{X}_1$ and $\bar{Z}_2$ act as products of $Z$ Paulis.

Lastly, we can branch $\bar{X}_1$ and $\bar{Z}_2$ to be on distinct leaves, or measure them in-place. The same goes for $\bar{Y}_3$.

Focussing first on $\bar{X}_1$,
\[\input{tikz_files/local_tannergraph_X1Z2_location.tikz}\]
we follow Ref.~\cite{williamson2024low} and construct a measurement graph,
\[\input{tikz_files/local_tannergraph_X1Z2_measure.tikz}\]
where new $Z$-checks are vertices in the measurement graph, new data qubits are edges in the measurement graph, and the new $X$-check gauge-fixes the cycle in the graph, cf. Def.~\ref{def:gauging_measurement}.

Repeating the process for $\bar{Z}_2$, we have:
\[\input{tikz_files/local_tannergraph_X1Z2_measure_pink.tikz}\]
where the bottom right $X$ check has two edges associated to it, as the $X$ check has weight 4 on the logical operator $\bar{Z}_2$, see Ref.~\cite[Rem.~4]{williamson2024low}.

Finally, for $\bar{Y}_3$ we have:
\[
\left[
\begin{array}{c|cccccccc|cccccccc}
\bar{Y}_3 & 1 & 0 & 0 & 0 & 1 & 0 & 0 & 0 &  0 & 0 & 0 & 1 & 1 & 0 & 1 & 1 \\
\end{array}
\right]
\]
which is also incident to all checks in the $\llbracket 8, 3, 3\rrbracket$ code; all of the new checks added to branch $\bar{X}_1$ and $\bar{Z}_2$ commute with the action of $\bar{Y}_3$ on each qubit, so $\bar{Y}_3$ is left undeformed by the prior branching. The local code, consisting of checks which anticommute with the action of $\bar{Y}_3$ on each qubit, is then,
\[\input{tikz_files/local_tannergraph_Y3.tikz}\]
As $\bar{X}_1$ and $\bar{Z}_2$ have been branched away from the support of $\bar{Y}_3$, we measure $\bar{Y}_3$ directly. To do so, permute the actions of $\bar{Y}_3$ on data qubits to be in the $Z$-basis, applying $H_1$ and $V_5$, where $V= HSH$ and $S = \sqrt{Z} = \begin{pmatrix}
    1 & 0 \\
    0 & i
\end{pmatrix}$. The permuted local code is,
\[\input{tikz_files/local_tannergraph_Y3_permuted.tikz}\]
where now $\bar{Y}_5$ acts as a product of $Z$ Paulis. Then insert the measurement graph as before,
\[\input{tikz_files/local_tannergraph_Y3_measure.tikz}\]
before finally permuting the bases back by applying $H_1$ and $V_5^\dagger$,
\[\input{tikz_files/local_tannergraph_Y3_permuteback.tikz}\]
where, to avoid clutter, we have abused notation and kept the new check measuring $\bar{Y}_5$ a $Z$-check, but with a $[1 | 1]$ below to indicate the edge which is actually measuring $Y$.

All told, the scheme for measuring $\bar{X}_1, \bar{Z}_2, \bar{Y}_3$ simultaneously used 27 new qubits and 30 new checks. This overhead is largely because the original code is already dense, so performing LDPC code surgery largely just adds overhead without preserving any useful property. In practice, measuring the $\llbracket 8, 3,3\rrbracket$ code could be done by adding 3 new checks, one for each of the logicals to be measured. 

One relevant saving, compared to the upper bounds on cost, is that because all logical qubits are measured, and all cycles in the measurement graphs are gauge-fixed, the deformed code will have $k=0$ and therefore $d=\infty$. As a consequence the measurement graphs do not need to be expanding to preserve code distance. All the other desiderata of Thm.~\ref{thm:desiderata} are satisfied. The same principle applies more generally: the more logical qubits are measured simultaneously on a quantum code, the fewer logical operators remain from the original code, so the less likely it is that cleaning them into the ancilla system will reduce the distance, see Ref.~\cite{williamson2024low,IGND}. As a consequence, the higher the rate of measurement, the lower the expansion of measurement graphs is likely to need to be. Additionally, the $X$-checks required to gauge-fix cycles are small enough that decongestion would not be required to maintain any sparsity, with the maximum weight of these checks being 3.

\section{Basis of Logical Operators for CSS Codes}\label{app:basis}

\CSSbasis*
\begin{proof}
    Let $m_Z$ and $m_X$ be the rank of $H_Z$ and $H_X$ respectively. Let us row-reduce the matrix \(H_Z\) to the form \(H_Z = [I_{m_Z}, R_Z]\), where \(I_{m_Z}\) denotes an identity matrix and \(R_Z\) is a \(m_Z\times (n-m_Z)\) matrix, discarding any all-zero rows.\footnote{This row echelon basis of stabilisers was studied in Ref.~\cite{CHNSZ} in the context of quantum code testing.}  
    Let \(L_Z\) be a \(k\times n\) matrix whose rows are a basis of $Z$ logical operators of the code \(Q\), and similar for \(L_X\).
    Stack \(H_Z\) with \(L_Z\) to form the following \((m_Z+k)\times n\) matrix:
    \[
    M_Z = \begin{bmatrix}
       I_{m_Z} & R_Z \\
       \multicolumn{2}{c}{L_Z}
    \end{bmatrix}
    \]
    Perform row-reduction on \(M_Z\), but with the following rule: we can add any of the top \(m_Z\) rows (corresponding to rows of \(H_Z\)) to any of the bottom \(k\) rows, and we can add any of the bottom \(k\) rows to any of the bottom \(k\) rows. We are not allowed to add any of the bottom \(k\) rows to any of the top \(m_Z\) rows.
    Up to permutation of columns, we can reduce \(M_Z\) to have the form
    \[
    M_Z = 
    \begin{bmatrix}
    I_{m_Z} & \multicolumn{2}{c}{R_Z} \\
    0_{k, m_Z} & I_k & L_Z'
    \end{bmatrix}
    \]
    where \(L_Z'\) is a \(k\times (n-m_Z-k)\) matrix, and \(0_{k,m_Z}\) denote a \(k\times m_Z\) matrix of zeros. Note that \(n-m_Z-k = m_X\).
    From here on, we will fix the ordering of columns and no longer permute them. 
    These last \(k\) rows are our \(Z\) logical operators \(z_i\).

    Now consider the matrix \(H_X\). We know that the rows of \(H_X\) span the dual space of the row space of \(M_Z\). 
    Therefore, let us write down a basis of this dual space, which will be an equivalent set of X-stabiliser generators of \(Q\).
    Let
    \[
    H_X' = \begin{bmatrix}
    \multicolumn{2}{c}{R_X} & I_{m_X}
    \end{bmatrix},
    \]
    where \(R_X\) is a \(m_X\times (n-m_X)\) matrix. 
    We need to find \(R_X\) such that every row \(r\) of \(H_X'\) satisfies \(M_Zr = 0\).

    Let us write \(R_Z = [R_{Z,1}, R_{Z,2}]\), where \(R_{Z,1}\) is a \(m_Z\times k\) matrix and \(R_{Z,2}\) is a \(m_Z\times m_X\) matrix. Then
    \[
    M_Z = 
    \begin{bmatrix}
    I_{m_Z} & R_{Z,1} & R_{Z,2} \\
    0_{k, m_Z} & I_k & L_Z'
    \end{bmatrix},
    \]
    and we will take 
    \[
    H_X' = \begin{bmatrix}
    (R_{Z,2})^\top + (R_{Z,1}L_Z')^\top & (L_Z')^\top & I_{m_X}
    \end{bmatrix}.
    \]
    One can check that \(M_ZH_X' = 0\), which means the rows of \(H_X'\) form a valid basis of X-stabilisers.
    We can now stack \(H_X'\) with \(L_X\), and get
    \[
    M_X = \begin{bmatrix}
       R_X & I_{m_X} \\
       \multicolumn{2}{c}{L_X}
    \end{bmatrix},
    \]
    and we apply stabilisers to operators such that
    \[
    M_X = \begin{bmatrix}
       \multicolumn{2}{c}{R_X} & I_{m_X} \\
       L_X' & J & 0_{k, m_X}
    \end{bmatrix},
    \]
    for some \(k\times m_Z\) matrix \(L_X'\) and \(k\times k\) matrix  \(J\).
    Note that the last \(k\) rows of \(M_X\) form a basis of X-logical operators of \(Q\).

    We claim that \(J\) must be full-rank. If not, there exists a vector \(x\) generated by the last \(k\) rows of \(M_X\) that has the form \(x = [x', 0, \cdots, 0]\) where \(x'\) is a length \(m_Z\) vector. 
    Consequently, \(H_Zx^\top = x'\), which contradicts the assumption that \(H_ZM_X^\top = 0\). 
    Therefore \(J\) must be full rank, we can reduce the last \(k\) rows (only adding these last \(k\) rows among themselves) of \(H_X\) to 
    \[
    \begin{bmatrix}
       L_X'' & I_k & 0_{k, m_X}
    \end{bmatrix}.
    \]
    These vectors are the \(X\) logical operators \(x_i\) of \(Q\), which satisfy the desired intersection rules with the \(Z\) logical operators.
\end{proof}

\section{Fault-tolerance proofs}\label{app:ft_proofs}

For all our proofs of fault-tolerance, we closely follow the approach in Ref.~\cite{WY}. For this it is necessary to recall some definitions and conventions from the literature. We follow the general approach to fault-tolerance via repeated syndrome measurements~\cite{De}, and the fault-distance is taken with respect to phenomenological noise; that is, there are faults on data qubits that can occur at any timestep, including initialisation and measurement, and there are measurement faults on syndrome qubits that can occur at any timestep.

The choice of phenomenological noise is justified as follows. Consider performing bare ancilla syndrome extraction, introducing a qubit for each check, entangling the incident data qubits, then measuring out the check qubit. When the spacetime code is LDPC, the syndrome extraction circuit fault-distance is lower-bounded by $d/c$, where $d$ is the phenomenological fault-distance of the spacetime code and $c$ is a constant. See, for example, the discussion in Ref.~\cite[Sec.~5]{gottesman2013fault}, or Ref.~\cite[Sec.~6.3]{he2025composable}. The constant $c$ is of vital importance in practice, and syndrome extraction circuits must be optimised to reduce it, but for the purposes of asymptotics it can be ignored. Moreover, as our parallel measurement scheme is generic, applicable to any stabiliser LDPC code, it is unfeasible to use circuit fault-distance directly, as we do not presume syndrome extraction circuits are defined in advance for the initial codes.

All our procedures begin at time $t_0$ followed by at least $d$ rounds of syndrome measurement in the initial distance-$d$ code $Q$. Then there is a code deformation step at time $t_i$ with $d$ rounds of syndrome measurement in the deformed code. When branching, we do not deform back immediately, but instead leave at least $d$ rounds of syndrome measurement before future gauging logical measurements are applied. Then finally the code is unbranched at time $t_o$.

As the fault-tolerance of gauging logical measurements was proved in Ref.~\cite{WY}, we do not prove it here, but instead take as given that following branching there is an intermediate series of rounds of syndrome measurement such that upon exit of the gauging logical measurements procedure we then unbranch the code, ready for the next timestep in the logical circuit. That is, we have the circuit
\[\input{tikz_files/branch_circuit.tikz}\]
where between each procedure and at each end there are $d$ syndrome measurement rounds, labelled `Idle'. These ``padding'' steps are required for the proofs to prevent timelike errors entering the spacetime in which each code deformation procedure is performed, however we expect they are not always necessary in general.\footnote{Regarding the idling steps at either end of the circuit, it is common in the study of fault-tolerant protocols to insert rounds of perfect syndrome extraction immediately before and after the protocol. However, this leads to a well-known composition problem~\cite{BHK}. When fault-tolerant protocols proved in this manner are performed sequentially they are no longer generally fault tolerant; cases where this can occur are described in Ref.~\cite{cowtan2025fast}. Inserting $d$ rounds of idling ensures that any nontrivial logical spacetime fault which extends from one protocol to another must have weight at least $d$, and so composition is preserved.} Indeed, we expect that the entire circuit above can be performed by initialising all new ancilla qubits, measuring all new checks for $d$ syndrome measurement rounds, and then measuring out all ancilla qubits, i.e.~using only approximately $d$ syndrome measurement rounds in total:
\[\input{tikz_files/branch_circuit2.tikz}\]
This more compact procedure would then be a single parallel gauging logical measurement, where the $Z$ checks used to infer a given $\bar{Z}_i$ logical measurement include those which clean an initial $v_i$ to its final location on the leaf of a branch. We do not prove that this is possible here, as such a proof would be cumbersome, and it does not affect the asymptotic time overhead.

We also follow Ref.~\cite{BHK} and use the convention that the initial and final rounds of measurement in the branching and unbranching procedures are perfect, which is justified by the ``padding'' as any fault that involves two code deformations, or both a code deformation and the initial or final boundary conditions, must have weight greater than $d$~\cite[Rem.~29]{WY}.

\begin{definition} [Space and time faults]

A space-fault is a Pauli error operator that occurs on some qubit during the implementation of a code deformation. A time-fault is a measurement error where the result of a
measurement is reported incorrectly during the implementation of a code deformation. A general spacetime
fault is a collection of space and time faults.
\end{definition}

We consider state initialisation faults to be time-faults that
are equivalent to space-faults by decomposing them into
a perfect initialisation followed by an error operator.

\begin{definition} [Detectors]
A detector is a collection of
state initialisations and measurements that yield a deterministic result, in the sense that the product of the
observed measurement results must be $+1$ independent
of the individual measurement outcomes, if there are no
faults in the procedure.
\end{definition}

\begin{remark}
    Some detectors have an empty collection of state initialisations, i.e. they consist only of measurements: for example, repeated measurement of a stabiliser check constitutes a detector. Other detectors include both state initialisations and measurements: for example, initialisation of a CSS code in the $\ket{0}^{\otimes n}$ state, followed by measurement of $Z$-type checks. In our case, we have both classes of detectors; the latter class arises from initialising new data qubits in the ancilla system then performing measurements upon them.
\end{remark}

\begin{definition} [Syndrome]

The syndrome caused by a
spacetime fault is defined to be the set of detectors that
are violated in the presence of the fault. That is, the
set of detectors which do not satisfy the constraint that
the observed measurement results multiply to +1 in the
presence of the fault.
\end{definition}

We follow Ref.~\cite{WY} in our convention for labelling time steps, whereby detectors and space errors occur at integer timesteps, but syndrome measurements and measurement errors occur at integer + half timesteps. Thus the first code deformation measurements are made at timestep $t_i + \frac{1}{2}$, and the final measurements before $d$ rounds of padding are made at timestep $t_o+\frac{1}{2}$.

Check $s_j$ is the $j$th check in the original code, and $\tilde{s}_j$ the (possibly deformed) check from the original code in the new deformed code.

We consider the ancillary system to be an incidence structure, i.e. a chain complex, called $\mathtt{C}$. While this incidence structure is not generally a cell complex, we nevertheless associate new $Z$ checks to vertices, new data qubits to hyperedges and new $X$ checks to faces; this is a labelling to match the similar proof in Ref.~\cite{williamson2024low}, where the incidence structure is strictly a cell complex.

We call a new $Z$ check in the branching procedure an $A_v$ check, corresponding to a vertex $v$. Similarly we have a $B_p$ check on a face $p$ in the $X$ basis at time $t$. 

$Z_e$ and $X_e$ act as $Z$ and $X$ Pauli operators on ancillary qubits associated with a hyperedge $e$. $X_i$ and $Z_i$ act as $X$ and $Z$ Pauli operators on a data qubit in the original code labelled $i$.

\subsection{Proof of Lemma~\ref{lem:branch_fault_tolerance}}\label{app:branching_ft}

Assume we are brute-force branching $\bar{Z}$ logicals. This does not lose generality, as the generalised branching from Lemma~\ref{lem:mixed_branching} is equivalent to branching $\bar{Z}$ logicals up to single-qubit Cliffords, using the same arguments as in Ref.~\cite{WY}. Before time $t_i$ we have the code $Q$, with distance $d$.
All new data qubits are initialised in the $\ket{+}$ state at time $t_i$, and all checks are measured from time $t_i+\frac{1}{2}$ to $t_o+\frac{1}{2}$. All new $X$ checks, in the absence of errors, deterministically return the +1 outcome. The new $Z$ checks, however, return random outcomes due to the $\ket{+}$ input states. This scenario is slightly different from the scenario of gauging logical measurements, as there we know that the product of the new $Z$ checks must yield the outcome of the logical measurement, in the absence of errors. Here we do not know the product of the outcomes. These outcomes dictate which Pauli frame \cite{Kn} we are in, and so we can reinterpret consistent $-1$ outcomes as $+1$ consistent outcomes; in other words, there are detectors between each successive $Z$ measurement, but not a detector given by the first $Z$ measurement and the initialisation of the new qubits in the $\ket{+}$ state. Hence these outcomes are very similar to the state preparation of $\ket{\bar{+}}^{\otimes k}$ in a CSS code: the $Z$ measurements induce a Pauli frame, which must then be consistent.

Throughout the discussion of the spacetime code, detectors (placed at integer timesteps) are always induced by at least one check. Therefore we use the notation that for a check $C$ at time $t+\frac{1}{2}$ or $t-\frac{1}{2}$ a corresponding detector at time $t$ which uses that check is labelled $C^t$.

\begin{lemma}[Spacetime code for branching]

The following form a generating set of the local detectors in the fault-tolerant branching procedure:

For $t < t_i$
\begin{itemize}
    \item $s_j^t$ which is given by repeated $s_j$ checks in the original code at times $t-\frac{1}{2}$, $t+\frac{1}{2}$.
\end{itemize}

For $t=t_i$
\begin{itemize}
    \item $B_p^t$ which is given by the measurement of $B_p$ at time $t_i+\frac{1}{2}$ together with the initialisation of new qubits in the $\ket{+}$ state at time $t_i-\frac{1}{2}$.
    \item $\tilde{s}_j^{t_i}$ which is given by the measurement of $s_j$, together with the initialisation of new qubits in the $\ket{+}$ state at time $t_i-\frac{1}{2}$, and the measurement of $\tilde{s}_j$ at time $t_i+\frac{1}{2}$.
\end{itemize}

For $t>t_i$
\begin{itemize}
    \item $A_v^t$ which is the repeated measurement of an $A_v$ check in the deformed code at times $t-\frac{1}{2}$ and $t+\frac{1}{2}$.
    \item $B_p^t$ which is the same for a $B_p$ check.
    \item $\tilde{s}_j^t$ which is the same for an $\tilde{s}_j$ check.
\end{itemize}

\end{lemma}
\proof
Away from the initial and final steps of the
code deformation, the brute-force branching procedure consists of repeatedly measuring the
same checks. In this setting any local detector must contain a pair of measurements of the same check. Any detector formed by
the measurement of the same check at times $(t,t+k)$ 
can be decomposed into detectors at time steps $(t,t+1),(t+1,t+2),...,(t+k-1,t+k)$. In other words, the time component of the spacetime code constitutes a repetition code~\cite{HDTV}.

During the code deformation there are also detectors involving the initialisation of new qubits. Due to measurement of $A_v$ terms during the code deformation, any detector involving the initialization of new qubits in the $X$ basis must include a collection of new qubits that corresponds to a product of $B_p$ and $\tilde{s}_j$ checks. Any such detector can be decomposed into those listed above which involve qubits at time $t_i$, combined with repeated measurement checks.

Local detectors that cross the deformation step at $t_i$ must include a check $s_j$ from the original code before or after the deformation, along with a deformed version $\tilde{s}_j$. Such a detector can be decomposed into repeated measurement detectors and one of the detectors involving new qubits listed above.
\endproof

Spacetime faults can be organised according to the spacetime syndromes they create:
\\

For $t < t_i$
\begin{itemize}
    \item An $s_j$ measurement fault at time $t+\frac{1}{2}$ violates the detectors $s_j^t$ and $s_j^{t + 1}$.
    \item A Pauli $Z_i$($X_i$) fault at time $t$ violates the $s_j^t$ detectors for all incident $s_j$ which do not commute with $Z_i$($X_i$).
\end{itemize}

For $t=t_i$
\begin{itemize}
    \item A Pauli $Z_i$($X_i$) operator fault at time $t$ violates the $\tilde{s}_j^t$ detectors for all $\tilde{s}_j$ which do not commute with $Z_i$($X_i$).
    \item A Pauli $Z_e$ fault at time $t$ violates the $B_p^t$ detectors for all $p$ incident to $e$ and the $\tilde{s}_j^t$ detectors for all $\tilde{s}_j$ checks which anticommute with $Z_e$.
    \item A $\ket{+}_e$ initialisation fault at time $t - \frac{1}{2}$ is equivalent to a Pauli $Z_e$ operator fault at time $t$ and so violates the same detectors.
    \item An $\tilde{s}_j$ fault at time $t+\frac{1}{2}$ violates the detectors $\tilde{s}_j^t$ and $\tilde{s}_j^{t+1}$.
\end{itemize}

For $t > t_i$
\begin{itemize}
    \item A Pauli $Z_i$ fault at time $t$ violates the $\tilde{s}_j^t$ detectors for all $\tilde{s}_j$ which do not commute with $Z_i$.
    \item A Pauli $X_i$ fault at time $t$ violates the
    $\tilde{s}_j^t$ detectors for all $\tilde{s}_j$ which do not commute with $X_i$, and the $A_v^t$ detectors for all $A_v$ which do not commute with $X_i$.
    \item A Pauli $X_e$ fault at time $t$ violates the $A_v^t$ detectors for all incident $A_v$ checks.
    \item A Pauli $Z_e$ operator fault at time $t$ violates the $B_p^t$ detectors for all incident $B_p$ checks, and $\tilde{s}_j$ detectors for all incident $\tilde{s}_j$ which anticommute with $Z_e$.
    \item An $\tilde{s}_j$ measurement fault at time $t+\frac{1}{2}$ violates the detectors $\tilde{s}_j^t$ and $\tilde{s}_j^{t + 1}$.
    \item An $A_v$ measurement fault at time $t+\frac{1}{2}$ violates the detectors $A_v^t$ and $A_v^{t+1}$.
    \item A $B_p$ measurement fault at time $t+\frac{1}{2}$ violates the detector $B_p^{t+1}$.
\end{itemize}

\begin{definition}[Spacetime logical fault]

    A spacetime logical fault is a collection of space and time faults that does not violate any detectors.
\end{definition}

\begin{definition}[Spacetime stabiliser]

    A spacetime stabiliser is a trivial spacetime logical fault in the sense that it is a collection of space and time faults that does not violate any detectors and does not anticommute with any logical faults.
\end{definition}

\begin{lemma}\label{lem:spacetime_stabs1}
    The following form a generating set of local spacetime stabilisers:

    For $t<t_i$
    \begin{itemize}
        \item A stabiliser check operator $s_j$ at time $t$.
        \item A pair of $X_i$ (or $Z_i$) faults at time $t$, $t+1$ together with measurement faults on all checks $s_j$ that do not commute with $X_i$ (or $Z_i$) at time $t+\frac{1}{2}$.
    \end{itemize}

    For $t=t_i$
    \begin{itemize}
        \item A stabiliser check operator $s_j$ or $X_e$ at time $t$.
        \item A pair of $Z_i$ faults at time $t, t+1$ along with measurement faults on all checks $\tilde{s}_j$ and $A_v$ that do not commute with $Z_i$ at time $t+1$.
        \item A pair of $X_i$ faults at time $t, t+1$ along with measurement faults on all checks $\tilde{s}_j$ that do not commute with $X_i$ at time $t+1$.
        \item A pair of $Z_e$ faults at times $t, t+1$, together with measurement faults on incident  $A_v$ checks and all incident $\tilde{s}_j$ checks that do not commute with $Z_e$ at time $t+\frac{1}{2}$.
        \item A $\ket{+}_e$ initialisation fault at time $t-\frac{1}{2}$ together with a $Z_e$ fault on that qubit at time $t$.
        \item An $X_e$ fault at time $t+1$ along with $A_v$ measurement faults for all incident $v$ at time $t+\frac{1}{2}$.
    \end{itemize}

    For $t>t_i$
    \begin{itemize}
        \item A stabiliser check operator $\tilde{s}_j$, $A_v$ or $B_p$, at time $t$.
        \item A pair of $Z_i$ faults at time $t, t+1$ along with measurement faults on all checks $\tilde{s}_j$ that do not commute with $Z_i$ at time $t+1$.
        \item A pair of $X_i$ faults at time $t, t+1$ along with measurement faults on all incident $A_v$ checks, and all checks $\tilde{s}_j$ that do not commute with $X_i$ at time $t+1$.
        \item A pair of $Z_e$ faults at time $t, t+1$ together with measurement faults on all incident $B_p$ checks and all incident $\tilde{s}_j$ checks that do not commute with $Z_e$ at time $t+1$.
        \item A pair of $X_e$ faults at time $t, t+1$, together with measurement faults on all incident $A_v$ checks at time $t+1$.
    \end{itemize}
\end{lemma}
\proof
First, recall from the organisation of spacetime faults that all initialisation errors are equivalent to Pauli operator errors. Then, any nonempty local spacetime stabiliser must involve a Pauli operator error, as otherwise the stabiliser would have to include all repeated measurement errors of some check.

A nonempty spacetime stabiliser containing a Pauli operator must either be a space stabiliser or contain a Pauli operator at another time. The product of the Pauli operators from
all time steps involved must itself be a space stabiliser. Any local spacetime stabiliser of this form can be constructed from a product of the spacetime stabilisers introduced above by first reconstructing the operators at the earliest time step at the cost of creating matching operators at the next time step, and so on until the final time step, where the product of the operators must now become trivial. This leaves a
local spacetime stabiliser with only measurement errors,
which must also be trivial. Hence, the original fault pattern is a product of the introduced spacetime stabilisers as claimed.
\endproof

\begin{definition}[Spacetime fault-distance]
    The spacetime fault-distance is the weight, counted in terms of single site Pauli errors and single measurement errors, of the minimal collection of faults that does not violate any detectors and is not a spacetime stabiliser.
\end{definition}

\begin{lemma}[Time fault-distance]
    The fault-distance for measurement and initialisation errors is at least $(t_o-t_i)$.
\end{lemma}
\proof
All pure measurement logical faults must start at the code deformation step at time $t_i$ and end at time $t_o$. Otherwise a measurement fault at time step $t$ must
be followed and preceded by another measurement fault
on the same type of check at time steps $t - 1, t + 1$. Thus a measurement and initialisation logical fault must have distance at least $(t_o-t_i)$.
\endproof

\begin{lemma}[Decoupling of space and time faults]
Any nontrivial spacetime logical fault is equivalent to the product of a
space logical fault and a time logical fault, up to multiplication with spacetime stabilisers.
\end{lemma}
\proof
Let $F$ be an arbitrary spacetime fault. The space component of $F$ can be cleaned into a chosen single timestep $t_i \leq t \leq t_o$ by multiplication with the spacetime stabilisers listed from Lemma~\ref{lem:spacetime_stabs1} that involve like Pauli operator faults at time steps $t, t+1$ along with the appropriate measurement faults. Therefore we can clean all the space component of $F$ to time $t_i$.

In the cleaned spacetime logical fault all measurement
errors must occur in the time steps $t_i \leq t \leq t_o$ as measurement errors outside this time window must propagate
to the initial or final time boundary, which has no measurement errors by assumption. The measurement faults now
form strings that propagate through time from $t_i+\frac{1}{2}$ to $t_o+\frac{1}{2}$, terminating at an $A_v$ or $B_p$ measurement fault, and so are time logical faults forming codewords in the repetition code timelike component of the spacetime code.
\endproof

\begin{lemma}[Spacetime fault-distance]\label{lem:spacetime_fd}
Brute-force branching is a fault-tolerant operation with fault-distance $d$, assuming we use at least $d$ rounds of stabiliser measurements.
\end{lemma}
\proof
Let $F$ be any spacetime logical fault that is not equivalent to a spacelike logical fault. Then $F$ must have support on all time steps $t+\frac{1}{2}$ such that $t_i\leq t \leq t_o$, and so if $t_o-t_i \geq d$ then their weight is bounded below by $d$.

Now let $F$ be a spacetime logical fault that is equivalent to a spacelike logical fault. We will clean $F$ to become a spacelike logical fault using the spacetime stabilisers. Observe that we can always clean $F$ to become a spacelike logical fault using the spacetime stabiliser generators from Lemma~\ref{lem:spacetime_stabs1} without increasing the weight, as each stabiliser preserves the parity of space and initialisation faults at a fixed location. That cleaned spacelike logical fault must have weight at least $d$, by Lemma~\ref{lem:branching_distance}. Therefore the weight of $F$ must be at least $d$.
\endproof

That unbranching after a gauging logical measurement is also a fault-tolerant operation can be deduced in a similar manner, where now every new data qubit is measured out in the $X$ basis at time $t = t_o$. Unbranching only uses one stabiliser check round of measurements, but we assume $d$ rounds of padding before and after, as previously. This gives a similar spacetime code.

\begin{lemma}[Spacetime code for unbranching]

The following form a generating set of the local detectors in the fault-tolerant unbranching procedure.

For $t < t_o$
\begin{itemize}
    \item $A_v^t$ which is the repeated measurement of an $A_v$ check in the deformed code at times $t-\frac{1}{2}$ and $t+\frac{1}{2}$.
    \item $B_p^t$ which is the same for a $B_p$ check.
    \item $\tilde{s}_j^t$ which is the same for an $\tilde{s}_j$ check.
\end{itemize}

For $t = t_o$
\begin{itemize}
    \item $B_p^t$ which is the measurement of $B_p$ at time $t_o-\frac{1}{2}$ along with the measurement of $X_e$ on each incident hyperedge data qubit at time $t_o+\frac{1}{2}$.
    \item $\tilde{s}_j^{t_o}$ which is the measurement of $\tilde{s}_j$ at time $t_o-\frac{1}{2}$ along with the measurement of $X_e$ on each incident hyperedge data qubit, and $s_j$ at time $t_o + \frac{1}{2}$.
\end{itemize}

For $t > t_o$
\begin{itemize}
    \item $s_j^t$ which is given by the collection of repeated $s_j$ checks in the original code at times $t-\frac{1}{2}$, $t-\frac{1}{2}$.
\end{itemize}

\end{lemma}

Observe that the spacetime detectors for $t < t_o$ are the same as for $t > t_i$ when branching. The same is true for spacetime faults
and spacetime stabilisers for $t < t_o$. Similarly, the spacetime detectors, faults and stabilisers for $t > t_o$ are the same as for $t < t_i$ when branching.

The spacetime faults at $t = t_o$, along with the spacetime syndromes they yield, are given by:
\begin{itemize}
    \item An $X_i$ (or $Z_i$) operator fault at time $t$ violates the $\tilde{s}_j^t$ detectors for all $\tilde{s}_j$ which do not commute with $X_i$ (or $Z_i$).
    \item A $Z_e$ operator fault on a new data qubit at time $t$ violates the $\tilde{s}_j^t$ and $B_p^t$ detectors for all $\tilde{s}_j$ and $B_p$ which do not commute with $Z$.
    \item An $X_e$-measurement read-out fault at time $t+\frac{1}{2}$ is equivalent to a $Z_e$ operator fault at time $t$ and violates the same detectors.
    \item An $s_j$ measurement fault at time $t+\frac{1}{2}$ violates the detectors $\tilde{s}_j^t$ and $s_j^{t+1}$.
    \item A $B_p$ measurement fault at time $t+\frac{1}{2}$ violates the detector $B_p^{t-1}$.
\end{itemize}

The spacetime stabilisers at $t=t_o$ have the generating set
\begin{itemize}
    \item A stabiliser check operator $\tilde{s}_j$, $A_v$ or $B_p$ at time $t$.
    \item A pair of $X_i$ ($Z_i$) operator faults at time $t$, $t+1$, together with measurement faults on all checks $s_j$ which do not commute with $X_i$ ($Z_i$).
    \item A Pauli $Z_e$ fault on a hyperedge data qubit at time $t$, along with an $X_e$-measurement read-out fault at time $t+\frac{1}{2}$.
    \item A Pauli $X_e$ fault on a hyperedge data qubit at time $t$.
    \item A Pauli $X_e$ error on a hyperedge data qubit at time $t-1$ along with measurement errors on incident $A_v$ checks at time $t-\frac{1}{2}$.
\end{itemize}

Together with the spacetime detectors, faults and stabilisers before and after the unbranching, one can then use the same arguments as in Lemma~\ref{lem:spacetime_fd} to conclude that unbranching has fault-distance $d$. Because of the `padding', we need only measure out the new data qubits and pre-existing stabilisers once. We argue that, as with branching and gauging logical measurements, this padding is typically unnecessary in practice.

\section{Simultaneous branching}\label{app:sim_branching}

Consider the case where we wish to perform $\bar{Z}$ and $\bar{X}$ measurements on $t$ different logical qubits in parallel, but the representatives we have chosen are overlapping. In this case, we can still brute-force branch the $\bar{Z}$ logicals, say, but the $\bar{X}$ logicals which overlapped with them will be deformed to have new support on the branch \cite[App.~F]{ZL}.

In particular, let $\{v_1,v_2,\cdots,v_{t_Z}\}$ be the $\bar{Z}$ logicals, and $V = \bigcup_i v_i$ be the union of data qubit support of these logicals.

\begin{lemma}
    Let each $\bar{X}$ logical $w_j$ to be measured commute with each $\bar{Z}$ logical contained in $V$. Then each $\bar{X}$ logical can be deformed such that it picks up support in the first layer of the $Z$-type tree.
\end{lemma}
\proof
There are two branches attached to $V$, one with support on $u_1$ and one on $u_2$, with $u_1$ and $u_2$ the unions of support for the first $\lceil \frac{t}{2} \rceil$ logicals and the last $\lfloor \frac{t}{2} \rfloor$ respectively. We have $u_1 \subset V$ and $u_2 \subset V$. For the first branch, let $\del_1$ be $H_X$ restricted to $u_1$. Then the first layer of the first branch is a $Z$ layer, with a code from data qubits to $Z$ checks given by $\del_1^\intercal$.
An overlapping $w_j$ will now anticommute with the $Z$ checks in the first layer in correspondence with $w_j \cap V$. As $w_j$ commutes with each logical in $V$, $w_j \cap u_1 \in \ker(\del_1)^\perp$, and hence $w_j \cap u_1$ is in the column space of $\del_1^\intercal$.
Therefore there is some set of new data qubits in the first layer such that extending $w_j$ to these data qubits, i.e. applying $X$ to those data qubits, turns off the $Z$ checks in bijection with $w_j \cap u_1$. As the next layer is an $X$ layer, there are no further anticommuting checks.

Repeating this procedure for the other branch gives a set of qubits in the first layer of that branch as well.
\endproof

We can see that, in the best case, $w_j$ only overlaps with $u_1$ and not $u_2$, so only new qubits in one branch are added to the support of $w_j$; in the worst case $w_j$ could have substantial overlap with both subsets, and hence needs support on both branches. In the best case, the number of new qubits required is equal to the number of checks $c$ which are applied to clean $w_j$ from $V$; in the worst case it could be $2c$.

Also observe that strictly speaking we do not need $w_j$ to commute with every $\bar{Z}$ logical in $V$: only each $\bar{Z}$ logical in $u_1$ and $u_2$. Therefore if the first layer of branching results in codes $\del_1$ and $\del_2$ which remove $\bar{Z}$ logicals which $w_j$ anticommuted with then the deformation to the first layer still applies.

\begin{lemma}
    Let each $\bar{X}$ logical $w_j$ to be measured commute with each $v_1,v_2\cdots,v_t$. Then each $\bar{X}$ logical can be deformed such that it picks up support in the $Z$-type tree.
\end{lemma}
\proof
In this case there are some anticommuting $\bar{Z}$ logicals contained in $V$, but $w_j$ still commutes with the logicals to be measured.
When $w_j$ commutes with every $\bar{Z}$ logical in $u_1$ and $u_2$ we can merely add support on the first layer as before.
When it does not, we can add support to the second layer of the first branch in bijection with $w_j \cap u_1$ to satisfy the $Z$ checks in the first layer. This then anticommutes with the $Z$ checks in the third layer. Continue until we have support extending through the tree, until at some layer every $\bar{Z}$ logical which anticommuted with $w_j$ has been removed. In the worst case we add the remaining support to $w_j$ on the last $Z$ layer, which is the penultimate layer of the tree, as the last layer is an $X$ layer with only each $v_i$ on leaves, and $[w_j,v_i]= 0$. 

Then repeat this procedure for the other branch with support on $u_2$.
\endproof

In the worst case the $\bar{X}$ logical $w_j$ can pick up support on $\CO(t\omega \log t)$ new qubits. This is because $w_j$ can pick up support on each level of the branching tree, with up to weight $\CO(t\omega)$ on each level, and there are $\CO(\log t)$ levels.

We can then branch the deformed $\bar{X}$ logicals to measure them separately, so in one logical Pauli measurement timestep we branch $\bar{Z}$ logicals and $\bar{X}$ logicals and then measure with gauging logical measurements. As the deformed $\bar{X}$ logicals have no support on the leaves, this $X$-type branching leaves the $\bar{Z}$ measurements unaffected. Because the procedure may increase the weights of $\bar{X}$ to be measured by $\CO(t\omega \log t)$, and the procedure gives no clear method to measure $\bar{Y}$ logical operators in parallel, we focus on the method of choosing a suitable basis and representatives for our logical operators which avoid overlap, thereby avoiding these problems.

In finite-size practical cases, the fact that the representatives to be measured could be much smaller initially than those given by Lemma~\ref{lem:rep_basis} could lead to branching with overlap being more efficient.

We argue that, as in Appendix~\ref{app:ft_proofs}, the branches can be initialised simultaneously, at the same time as the data qubits required for gauging measurements, and the whole procedure performed with $d$ timesteps in total while maintaining fault-distance $d$, but we do not prove this here.

%% file: tikz_files/branch_circuit2.tikz
\begin{tikzpicture}
	\begin{pgfonlayer}{nodelayer}
		\node [style=none] (0) at (-3.5, 0) {};
		\node [style=none] (1) at (3.5, 0) {};
		\node [style=gate] (3) at (0, 0) {Measure};
	\end{pgfonlayer}
	\begin{pgfonlayer}{edgelayer}
		\draw (0.center) to (1.center);
	\end{pgfonlayer}
\end{tikzpicture}

%% file: parallel_surgery_PRXQresub.bbl
\begin{thebibliography}{10}

\bibitem{google2023suppressing}
Google~Quantum AI.
\newblock {Suppressing quantum errors by scaling a surface code logical qubit}.
\newblock {\em Nature}, 614(7949):676--681, 2023.

\bibitem{acharya2024quantumerrorcorrectionsurface}
Google~Quantum AI and Collaborators.
\newblock {Quantum error correction below the surface code threshold}.
\newblock {\em Nature}, 638:920--926, 2024.

\bibitem{bluvstein2024logical}
D.~Bluvstein, S.~J. Evered, A.~A. Geim, S.~H. Li, H.~Zhou, T.~Manovitz, et~al.
\newblock {Logical quantum processor based on reconfigurable atom arrays}.
\newblock {\em Nature}, 626(7997):58--65, 2024.
\newblock doi:10.1038/s41586-023-06927-3.

\bibitem{rodriguez2024experimental}
P.~S. Rodriguez, J.~M. Robinson, P.~N. Jepsen, Z.~He, C.~Duckering, C.~Zhao, K.~Wu, J.~Campo, K.~Bagnall, M.~Kwon, et~al.
\newblock {Experimental demonstration of logical magic state distillation}.
\newblock {\em arXiv preprint arXiv:2412.15165}, 2024.

\bibitem{lacroix2024scaling}
N.~Lacroix, A.~Bourassa, F.~J.~H. Heras, L.~M. Zhang, et~al.
\newblock {Scaling and logic in the color code on a superconducting quantum processor}.
\newblock {\em Nature}, 2025.

\bibitem{BE2}
N.~P. Breuckmann and J.~N. Eberhardt.
\newblock {Quantum Low-Density Parity-Check Codes}.
\newblock {\em PRX Quantum}, 2(4):040101, 2021.
\newblock doi:10.1103/PRXQuantum.2.040101.

\bibitem{BCGMRY}
S.~Bravyi, A.~W. Cross, J.~M. Gambetta, D.~Maslov, P.~Rall, and T.~J. Yoder.
\newblock {High-threshold and low-overhead fault-tolerant quantum memory}.
\newblock {\em Nature}, 627:778--782, 2024.
\newblock doi:10.1038/s41586-024-07107-7.

\bibitem{XBAP}
Q.~Xu, J.~P.~B. Ataides, C.~A. Pattison, N.~Raveendran, D.~Bluvstein, et~al.
\newblock {Constant-overhead fault-tolerant quantum computation with reconfigurable atom arrays}.
\newblock {\em Nature Physics}, 20(7):1084--1090, 2024.
\newblock doi:10.1038/s41567-024-02479-z.

\bibitem{Kit}
A.~Kitaev.
\newblock {Fault-tolerant quantum computation by anyons}.
\newblock {\em Ann. Phys.}, 303:3--20, 2003.
\newblock doi:10.1016/S0003-4916\%2802\%2900018-0.

\bibitem{HFDM}
D.~Horsman, A.~G. Fowler, S.~Devitt, and R.~Van Meter.
\newblock {Surface code quantum computing by lattice surgery}.
\newblock {\em New J. Phys.}, 14:123011, 2012.
\newblock doi:10.1088/1367-2630/14/12/123011.

\bibitem{moussa2016transversal}
J.~Moussa.
\newblock {Transversal Clifford gates on folded surface codes}.
\newblock {\em Physical Review A}, 94:042316, 2016.
\newblock doi:10.1103/PhysRevA.94.042316.

\bibitem{gidney2024magic}
C.~Gidney, N.~Shutty, and C.~Jones.
\newblock {Magic state cultivation: growing T states as cheap as CNOT gates}.
\newblock doi:10.48550/arXiv.2409.17595.

\bibitem{grassl2013leveraging}
Markus Grassl and Martin Roetteler.
\newblock {Leveraging automorphisms of quantum codes for fault-tolerant quantum computation}.
\newblock In {\em 2013 IEEE International Symposium on Information Theory}, pages 534--538. IEEE, 2013.

\bibitem{Burt1}
N.~P. Breuckmann and S.~Burton.
\newblock {Fold-Transversal Clifford Gates for Quantum Codes}.
\newblock {\em Quantum}, 8:1372, 2024.
\newblock doi:10.22331/q-2024-06-13-1372.

\bibitem{quintavalle2023partitioning}
A.~Quintavalle, P.~Webster, and M.~Vasmer.
\newblock {Partitioning qubits in hypergraph product codes to implement logical gates}.
\newblock {\em Quantum}, 7:1153, 2023.
\newblock doi:10.22331/q-2023-10-24-1153.

\bibitem{ES}
J.~N. Eberhardt and V.~Steffan.
\newblock {Logical Operators and Fold-Transversal Gates of Bivariate Bicycle Codes}.
\newblock {\em IEEE Transactions on Information Theory}, 71(2):1140--1152, 2025.

\bibitem{sayginel2024fault}
Hasan Sayginel, Stergios Koutsioumpas, Mark Webster, Abhishek Rajput, and Dan~E Browne.
\newblock {Fault-Tolerant Logical Clifford Gates from Code Automorphisms}.
\newblock {\em arXiv preprint arXiv:2409.18175}, 2024.

\bibitem{malcolm2025computing}
A.~Malcolm, A.~Glaudell, P.~Fuentes, D.~Chandra, et~al.
\newblock {Computing Efficiently in QLDPC Codes}.
\newblock doi:10.48550/arXiv.2502.07150.

\bibitem{zhu2023non}
Guanyu Zhu, Shehryar Sikander, Elia Portnoy, Andrew~W Cross, and Benjamin~J Brown.
\newblock {Non-{Clifford} and parallelizable fault-tolerant logical gates on constant and almost-constant rate homological quantum {LDPC} codes via higher symmetries}.
\newblock {\em arXiv preprint arXiv:2310.16982}, 2023.

\bibitem{scruby2024quantum}
Thomas~R Scruby, Arthur Pesah, and Mark Webster.
\newblock {Quantum rainbow codes}.
\newblock {\em arXiv preprint arXiv:2408.13130}, 2024.

\bibitem{golowich2024quantum}
Louis Golowich and Ting-Chun Lin.
\newblock {Quantum {LDPC} Codes with Transversal Non-{Clifford} Gates via Products of Algebraic Codes}.
\newblock {\em arXiv preprint arXiv:2410.14662}, 2024.

\bibitem{lin2024transversal}
Ting-Chun Lin.
\newblock {Transversal non-{Clifford} gates for quantum {LDPC} codes on sheaves}.
\newblock {\em arXiv preprint arXiv:2410.14631}, 2024.

\bibitem{hsin2024classifying}
Po-Shen Hsin, Ryohei Kobayashi, and Guanyu Zhu.
\newblock {Classifying Logical Gates in Quantum Codes via Cohomology Operations and Symmetry}.
\newblock {\em arXiv preprint arXiv:2411.15848}, 2024.

\bibitem{breuckmann2024cups}
Nikolas~P Breuckmann, Margarita Davydova, Jens~N Eberhardt, and Nathanan Tantivasadakarn.
\newblock {Cups and Gates {I}: Cohomology invariants and logical quantum operations}.
\newblock {\em arXiv preprint arXiv:2410.16250}, 2024.

\bibitem{zhu2025topological}
Guanyu Zhu.
\newblock {A topological theory for {qLDPC}: non-{Clifford} gates and magic state fountain on homological product codes with constant rate and beyond the ${N^{1/3}}$ distance barrier}.
\newblock {\em arXiv preprint arxiv:2501.19375}, 2025.

\bibitem{HJY}
S.~Huang, T.~Jochym-O'Connor, and T.~J. Yoder.
\newblock {Homomorphic Logical Measurements}.
\newblock {\em PRX Quantum}, 4:030301, 2023.
\newblock doi:10.1103/PRXQuantum.4.030301.

\bibitem{XZ}
Q.~Xu, H.~Zhou, G.~Zheng, D.~Bluvstein, J.~P.~B. Ataides, M.~D. Lukin, and L.~Jiang.
\newblock {Fast and Parallelizable Logical Computation with Homological Product Codes}.
\newblock {\em Phys. Rev. X}, 15:021065, May 2025.

\bibitem{Coh}
L.~Z. Cohen, I.~H. Kim, S.~D. Bartlett, and B.~J. Brown.
\newblock {Low-overhead fault-tolerant quantum computing using long-range connectivity}.
\newblock {\em Sci. Adv.}, 8, eabn1717, 2022.
\newblock doi:10.1126/sciadv.abn1717.

\bibitem{De}
E.~Dennis, A.~Kitaev, A.~Landahl, and J.~Preskill.
\newblock {Topological quantum memory}.
\newblock {\em J. Math. Phys.}, 43(9):4452--4505, September 2002.
\newblock doi:10.1063/1.1499754.

\bibitem{ZZ}
H.~Zhou et~al.
\newblock {Algorithmic Fault Tolerance for Fast Quantum Computing}.
\newblock doi:10.48550/arXiv.2406.17653.

\bibitem{EK}
B.~Eastin and E.~Knill.
\newblock {Restrictions on Transversal Encoded Quantum Gate Sets}.
\newblock {\em Phys. Rev. Lett.}, 102:110502, 2009.
\newblock doi:10.1103/PhysRevLett.102.110502.

\bibitem{he2025quantum}
Z.~He, V.~Vaikuntanathan, A.~Wills, and R.~Y. Zhang.
\newblock {Quantum Codes with Addressable and Transversal Non-Clifford Gates}.
\newblock {\em arXiv preprint arXiv:2502.01864}, 2025.

\bibitem{BSS}
S.~Bravyi, G.~Smith, and J.~A. Smolin.
\newblock {Trading Classical and Quantum Computational Resources}.
\newblock {\em Phys. Rev. X}, 6:021043, 2016.
\newblock doi:10.1103/PhysRevX.6.021043.

\bibitem{litinski2019game}
D.~Litinski.
\newblock {A Game of Surface Codes: Large-Scale Quantum Computing with Lattice Surgery}.
\newblock {\em Quantum}, 3 p (2019,3):128, 2019.
\newblock doi:10.22331/q-2019-03-05-128.

\bibitem{VLCABT}
C.~Vuillot, L.~Lao, B.~Criger, C.~G. Almudéver, K.~Bertels, and B.~M. Terhal.
\newblock {Code deformation and lattice surgery are gauge fixing}.
\newblock {\em New J. Phys.}, 21:033028, 2019.
\newblock doi:10.1088/1367-2630/ab0199.

\bibitem{hastings2016weight}
M.~B Hastings.
\newblock {Weight reduction for quantum codes}.
\newblock {\em arXiv preprint arXiv:1611.03790}, 2016.

\bibitem{hastings2021quantum}
M.~B Hastings.
\newblock {On quantum weight reduction}.
\newblock {\em arXiv preprint arXiv:2102.10030}, 2021.

\bibitem{Sabo2024}
E.~Sabo, L.~G. Gunderman, B.~Ide, M.~Vasmer, and G.~Dauphinais.
\newblock {Weight-Reduced Stabilizer Codes with Lower Overhead}.
\newblock {\em PRX Quantum}, 5:040302, Oct 2024.

\bibitem{WY}
D.~J. Williamson and T.~J. Yoder.
\newblock {Low-overhead fault-tolerant quantum computation by gauging logical operators}.
\newblock doi:10.48550/arXiv.2410.02213.

\bibitem{IGND}
B.~Ide, M.~G. Gowda, P.~J. Nadkarni, and G.~Dauphinais.
\newblock {Fault-Tolerant Logical Measurements via Homological Measurement}.
\newblock {\em Phys. Rev. X}, 15:021088, Jun 2025.

\bibitem{CowBu}
A.~Cowtan and S.~Burton.
\newblock {CSS code surgery as a universal construction}.
\newblock {\em Quantum}, 8:1344, 2024.
\newblock doi:10.22331/q-2024-05-14-1344.

\bibitem{Cow24}
A.~Cowtan.
\newblock {{SSIP}: automated surgery with quantum {LDPC} codes}.
\newblock {\em arXiv:2407.09423}, 2024.

\bibitem{CHRY}
A.~Cross, Z.~He, P.~Rall, and T.~J. Yoder.
\newblock {Improved QLDPC Surgery: Logical Measurements and Bridging Codes}.
\newblock {\em arXiv preprint arXiv:2407.18393}, 2024.

\bibitem{FH}
M.~Freedman and M.~Hastings.
\newblock {Building manifolds from quantum codes}.
\newblock {\em Geometric and Functional Analysis}, 31:855, 2021.
\newblock doi:10.1007/s00039-021-00567-3.

\bibitem{SJY}
E.~Swaroop, T.~Jochym-O'Connor, and T.~J. Yoder.
\newblock {Universal adapters between quantum LDPC codes}.
\newblock doi:10.48550/arXiv.2410.03628.

\bibitem{ZL}
Guo Zhang and Ying Li.
\newblock {Time-Efficient Logical Operations on Quantum Low-Density Parity Check Codes}.
\newblock {\em Phys. Rev. Lett.}, 134:070602, Feb 2025.

\bibitem{breuckmann2017hyperbolic}
Nikolas~P Breuckmann, Christophe Vuillot, Earl Campbell, Anirudh Krishna, and Barbara~M Terhal.
\newblock {Hyperbolic and semi-hyperbolic surface codes for quantum storage}.
\newblock {\em Quantum Science and Technology}, 2(3):035007, 2017.

\bibitem{CC}
C.~Chamberland and E.~T. Campbell.
\newblock {Universal Quantum Computing with Twist-Free and Temporally Encoded Lattice Surgery}.
\newblock {\em PRX Quantum}, 3:010331, 2022.
\newblock doi:10.1103/PRXQuantum.3.010331.

\bibitem{gottesman1997stabilizer}
Daniel Gottesman.
\newblock {\em {Stabilizer codes and quantum error correction}}.
\newblock PhD thesis, California Institute of Technology, 1997.

\bibitem{williamson2024low}
Dominic~J Williamson and Theodore~J Yoder.
\newblock {Low-overhead fault-tolerant quantum computation by gauging logical operators}.
\newblock {\em arXiv preprint arXiv:2410.02213}, 2024.

\bibitem{swaroop2024universal}
Esha Swaroop, Tomas Jochym-O'Connor, and Theodore~J Yoder.
\newblock {Universal adapters between quantum {LDPC} codes}.
\newblock {\em arXiv preprint arXiv:2410.03628}, 2024.

\bibitem{cross2024improved}
Andrew Cross, Zhiyang He, Patrick Rall, and Theodore Yoder.
\newblock {Improved Q{LDPC} Surgery: Logical Measurements and Bridging Codes}.
\newblock {\em arXiv preprint arXiv:2407.18393}, 2024.

\bibitem{ide2024fault}
Benjamin Ide, Manoj~G. Gowda, Priya~J. Nadkarni, and Guillaume Dauphinais.
\newblock Fault-tolerant logical measurements via homological measurement.
\newblock {\em Phys. Rev. X}, 15:021088, Jun 2025.

\bibitem{HZW}
A.~Hamma, P.~Zanardi, and X.~G. Wen.
\newblock {String and membrane condensation on three-dimensional lattices}.
\newblock {\em Phys. Rev. B}, 72:035307, 2005.
\newblock doi:10.1103/PhysRevB.72.035307.

\bibitem{PK}
P.~Panteleev and G.~Kalachev.
\newblock {Asymptotically good Quantum and locally testable classical LDPC codes}.
\newblock In {\em Proceedings of the 54th Annual ACM SIGACT Symposium on Theory of Computing}, STOC 2022, page 375–388, New York, NY, USA, 2022. Association for Computing Machinery.

\bibitem{he2025extractors}
Z.~He, A.~Cowtan, D.~J. Williamson, and T.~J. Yoder.
\newblock {Extractors: QLDPC Architectures for Efficient Pauli-Based Computation}.
\newblock {\em arXiv preprint arXiv:2503.10390}, 2025.

\bibitem{yoder2025tour}
T.~J. Yoder, E.~Schoute, P.~Rall, E.~Pritchett, J.~M. Gambetta, A.~W. Cross, M.~Carroll, and M.~E. Beverland.
\newblock {Tour de gross: A modular quantum computer based on bivariate bicycle codes}.
\newblock {\em arXiv:2506.03094}, 2025.

\bibitem{CHP}
T.~Carette, D.~Horsman, and S.~Perdrix.
\newblock {SZX-Calculus: Scalable Graphical Quantum Reasoning, 44th International Symposium on Mathematical Foundations of Computer Science (MFCS 2019)}.
\newblock doi:10.4230/LIPIcs.MFCS.2019.55.

\bibitem{TZ}
J.~P. Tillich and G.~Z{\'e}mor.
\newblock {Quantum LDPC Codes With Positive Rate and Minimum Distance Proportional to the Square Root of the Blocklength}.
\newblock {\em IEEE Transactions on Information Theory}, 60(2):1193--1202, February 2014.
\newblock doi:10.1109/TIT.2013.2292061.

\bibitem{BT}
S.~Bravyi and B.~Terhal.
\newblock {A no-go theorem for a two-dimensional self-correcting quantum memory based on stabilizer codes}.
\newblock {\em New J. Phys.}, 11:043029, 2009.
\newblock doi:10.1088/1367-2630/11/4/043029.

\bibitem{tjoc2025basis}
Tomas Jochym-O'Connor.
\newblock {Personal communications}.

\bibitem{shi2025stabilizer}
Yu~Shi, Ashlesha Patil, and Saikat Guha.
\newblock {Stabilizer Entanglement Distillation and Efficient Fault-Tolerant Encoders}.
\newblock {\em PRX Quantum}, 6(1), March 2025.

\bibitem{aasen2025geometrically}
David Aasen, Jeongwan Haah, Matthew~B Hastings, and Zhenghan Wang.
\newblock {Geometrically Enhanced Topological Quantum Codes}.
\newblock {\em arXiv preprint arXiv:2505.10403}, 2025.

\bibitem{vardy2002intractability}
Alexander Vardy.
\newblock The intractability of computing the minimum distance of a code.
\newblock {\em IEEE Transactions on Information Theory}, 43(6):1757--1766, 2002.

\bibitem{dumer2003hardness}
I.~Dumer, D.~Micciancio, and M.~Sudan.
\newblock Hardness of approximating the minimum distance of a linear code.
\newblock {\em IEEE Transactions on Information Theory}, 49(1):22--37, 2003.

\bibitem{KLL}
I.~H. Kim, Y.~Liu, S.~Pallister, W.~Pol, S.~Roberts, and E.~Lee.
\newblock {Fault-tolerant resource estimate for quantum chemical simulations: Case study on Li-ion battery electrolyte molecules}.
\newblock {\em Phys. Rev. Res.}, 4:023019, Apr 2022.

\bibitem{hsieh2025simplified}
Min-Hsiu Hsieh, Xingjian Li, and Ting-Chun Lin.
\newblock Simplified quantum weight reduction with optimal bounds.
\newblock {\em arXiv preprint arXiv:2510.09601}, 2025.

\bibitem{zheng2025highrate}
Guo Zheng, Liang Jiang, and Qian Xu.
\newblock High-rate surgery: towards constant-overhead logical operations.
\newblock {\em arXiv preprint arXiv:2510.08523}, 2025.

\bibitem{HDTV}
T.~Hillmann, G.~Dauphinais, I.~Tzitrin, and M.~Vasmer.
\newblock {Single-shot and measurement-based quantum error correction via fault complexes}.
\newblock doi:10.48550/arXiv.2410.12963.

\bibitem{baspin2025fast}
Nouédyn Baspin, Lucas Berent, and Lawrence~Z. Cohen.
\newblock Fast surgery for quantum ldpc codes, 2025.

\bibitem{cowtan2025fast}
Alexander Cowtan, Zhiyang He, Dominic~J. Williamson, and Theodore~J. Yoder.
\newblock Fast and fault-tolerant logical measurements: Auxiliary hypergraphs and transversal surgery.
\newblock {\em arXiv preprint arXiv:2510.14895}, 2025.

\bibitem{TkZ}
{TikZit}.
\newblock https://tikzit.github.io/index.html.

\bibitem{Mac}
S.~MacLane.
\newblock {\em {Categories for the Working Mathematician}}.
\newblock Graduate Texts in Mathematics 5. Springer, second edition, 1997.
\newblock doi:10.1007/978-1-4757-4721-8.

\bibitem{AC}
B.~Audoux and A.~Couvreur.
\newblock {On tensor products of CSS Codes}.
\newblock {\em Ann. Inst. Henri Poincaré Comb. Phys. Interact.}, 6(2):239--287, 2019.
\newblock doi:10.4171/aihpd/71.

\bibitem{Weib}
C.~A. Weibel.
\newblock {\em {An Introduction to Homological Algebra}}.
\newblock Cambridge Studies in Advanced Mathematics, Cambridge University Press, 1994.
\newblock doi:10.1017/CBO9781139644136.

\bibitem{cowtan2025homology}
Alexander Cowtan.
\newblock Homology, hopf algebras and quantum code surgery.
\newblock {\em arXiv preprint arXiv:2508.01496}, 2025.

\bibitem{pesah2025fault}
Arthur Pesah, Austin~K. Daniel, Ilan Tzitrin, and Michael Vasmer.
\newblock Fault-tolerant transformations of spacetime codes, 2025.

\bibitem{CHNSZ}
A.~Cross, Z.~He, A.~Natarajan, M.~Szegedy, and G.~Zhu.
\newblock {Quantum Locally Testable Code with Constant Soundness}.
\newblock {\em Quantum}, 8:1501, 2024.
\newblock doi:10.22331/q-2024-10-18-1501. p (2024,10).

\bibitem{gottesman2013fault}
Daniel Gottesman.
\newblock {Fault-tolerant quantum computation with constant overhead}.
\newblock {\em arXiv preprint arXiv:1310.2984}, 2013.

\bibitem{he2025composable}
Zhiyang He, Quynh~T. Nguyen, and Christopher~A. Pattison.
\newblock Composable quantum fault-tolerance.
\newblock {\em arXiv preprint arXiv:2508.08246}, 2025.

\bibitem{BHK}
M.~E. Beverland, S.~Huang, and V.~Kliuchnikov.
\newblock {Fault tolerance of stabilizer channels}.
\newblock doi:10.48550/arXiv.2401.12017.

\bibitem{Kn}
E.~Knill.
\newblock {Quantum computing with realistically noisy devices}.
\newblock {\em Nature}, 434(7029):39--44, 2005.
\newblock doi:10.1038/nature03350.

\end{thebibliography}
